\documentclass[two-column,jgrga]{AGUTeX}
\newcommand{\planss}{Planetary Space Science}
\newcommand{\nat}{Nature}

  \usepackage[pdftex]{graphicx}
  \usepackage{multirow}
  \usepackage{amssymb}
  \usepackage{lscape}
  \usepackage{graphics}
\usepackage{epic}

\authorrunninghead{FLESHMAN ET AL.}

\titlerunninghead{ENCELADUS TORUS MODEL}

\authoraddr{Bobby Fleshman,
Homer L. Dodge Department of Physics and Astronomy, University of Oklahoma, 
440 W. Brooks, Norman, Oklahoma 73019, USA
(fleshman@nhn.ou.edu)}

\begin{document}

%
%

\title{A Sensitivity Study of the Enceladus Torus}
%

%
%


\author{B. L. Fleshman}
\affil{Laboratory for Atmospheric and Space Physics,
University of Colorado, Boulder, Colorado, USA\\
Homer L. Dodge Department of Physics and Astronomy, University of Oklahoma, 
Norman, Oklahoma, USA}
\author{P. A. Delamere}
\affil{Laboratory for Atmospheric and Space Physics,
University of Colorado, Boulder, Colorado, USA}
\author{F. Bagenal}
\affil{Laboratory for Atmospheric and Space Physics,
University of Colorado, Boulder, Colorado, USA}



%
%
%

%
%


\begin{abstract}
We have developed a homogeneous model of physical chemistry to investigate the neutral-dominated, water-based Enceladus torus.  Electrons are treated as the summation of two isotropic Maxwellian distributions$-$a thermal component and a hot component.  The effects of electron impact, electron recombination, charge exchange, and photochemistry are included.  The mass source is neutral H$_2$O, and a rigidly-corotating magnetosphere introduces energy via pickup of freshly-ionized neutrals.  A small fraction of energy is also input by Coulomb collisions with a small population ($<$\,1\%) of supra-thermal electrons.  Mass and energy are lost due to radial diffusion, escaping fast neutrals produced by charge exchange and recombination, and a small amount of radiative cooling.  We explore a constrained parameter space spanned by water source rate, ion radial diffusion, hot-electron temperature, and hot-electron density.  The key findings are: (1) radial transport must take longer than 12 days; (2) water is input at a rate of 100--180 kg s$^{-1}$; (3) hot electrons have energies between 100 and 250 eV; (4) neutrals dominate ions by a ratio of 40:1 and continue to dominate even when thermal electrons have temperatures as high as $\approx$\,5 eV; (5) hot electrons do not exceed 1\% of the total electron population within the torus; (6) if hot electrons alone drive the observed longitudinal variation in thermal electron density, then they also drive a significant variation in ion composition.
\end{abstract}

%
%

%

\begin{article}

%
%

\section{Introduction}
Absorption of UV starlight during occultation of Saturn's moon Enceladus showed that it continuously ejects neutral H$_2$O at a rate of $\approx$\,150--300\,kg\,s$^{-1}$ from water-ice geysers located at its southern pole \citep{2006Sci...311.1422H}.  Models suggest that the water and its chemical by-products form an extended neutral-dominated torus centered on the orbit of Enceladus \citep{jurac2005}.  Similarly, Jupiter's volcanic moon Io emits a mixture of SO$_2$ and S$_2$ at a rate of $\approx$\,1\,ton\,s$^{-1}$, and chemical by-products produce a plasma torus centered on the orbit of Io (see review by \cite{thomas04}).  The Hubble Space Telescope (HST) observations by \cite{1993Natur.363..329S} revealed that the neutral-to-ion ratio in the Enceladus torus ($\approx$\,10) is three orders of magnitude greater than in the Io torus ($\approx$\,10$^{-2}$).  Compositional differences and the degree of ionization within these two systems can be attributed to their chemistry (Io's based on sulfur dioxide and Enceladus's based on water) as well as the fact that fresh ions are picked with five times more energy in the Io torus than in the Enceladus torus \citep{2007GeoRL..3409105D}.  

An important lesson learned from studying the physical chemistry of the Io torus is that a small fraction of hot electrons ($<$\,1\%) play a critical role in determining composition.  To model the Cassini UltraViolet Imaging Spectrograph (UVIS) data obtained during Cassini's E2 flyby of Jupiter (October 2000 to March 2001), \cite{steffl2,steffl1,steffl3,steffl4} adapted the \cite{2003JGRA..108.1276D} Io torus model.  \textit{Steffl et al.}\ used their models to study radial, temporal, and azimuthal variation in mixing ratios (ion-to-electron density ratios), thermal-electron density, and thermal-electron temperature.  \cite{steffl4} concluded that hot electrons are necessary for the Io torus energy budget and that two modulations of the hot-electron population are required to reproduce both the temporal and spatial variations in composition observed in the data, one modulating in Jupiter's System III longitude, the other in System IV.  We anticipate that hot electrons are similarly important in Saturn's Enceladus torus.  

\cite{2007GeoRL..3409105D} developed a simplified oxygen-based model to compare the Enceladus and Io tori.  They found that collisional heating by a population of hot electrons is much less important at Enceladus, contributing only 0.5\% of the energy to the torus, compared to 60\% at Io.  They also cited two major reasons for the discrepancy in the neutral-to-ion ratio between the two systems.  First, newly created ions are picked up in the Io torus by Jupiter's magnetosphere at roughly five times the energies as are those in the Enceladus torus by Saturn's magnetosphere.  The higher-energy pickup ions in the Io torus warm the thermal electrons, which then reduces the neutral-to-ion ratio via impact ionization.  Second, because of the high abundance of molecular ions compared with atomic oxygen ions in the Enceladus torus (e.g., \cite{sittler2005}), \cite{2007GeoRL..3409105D} expected that molecular dissociative recombination (not included in their model) will therefore drive the ratio even higher in the Enceladus torus.  From the conclusion of \cite{2007GeoRL..3409105D}:  ``The addition of the full water-group molecular chemistry will introduce an additional plasma sink through dissociative recombination of the molecular ions.  Therefore, our simplified O-based chemistry likely represents a lower limit for the neutral/ion ratio.'' To investigate the consequences of a water-based Enceladus torus dominated by molecular chemistry, we have improved on the Delamere model by including a comprehensive set of water-based reactions and species to more accurately estimate steady-state densities and temperatures of ions and electrons.  We also add neutral and ionized molecules.  Molecules are more abundant in the Enceladus torus where low plasma density allows H$_2$O to escape from Enceladus largely intact, whereas the more energetic local plasma interaction at Io results in dissociation of SO$_2$ \citep{2008JGRA..11309208D}.  Moreover, thermal electrons ($\approx$\,2\,eV) throughout the Enceladus torus dissociate H$_2$O approximately ten times less easily than thermal electrons ($\approx$\,5\,eV) in the Io torus dissociate SO$_2$ (\textit{V.\ Dols, personal comm.}).

Previous models of molecular chemistry in Saturn's magnetosphere were driven by Pioneer 11 and Voyager 1 and 2 observations \citep{frank1980,trainor1980,wolfe1980,bridge1981,bridge1982,sittler1983}.  \cite{richardson1986} showed the importance of recombination under conditions of slow radial transport.  Using essentially the same chemical reactions as \cite{richardson1986}, \cite{1998JGR...10320245R} determined ion and neutral lifetimes within Saturn's inner magnetosphere ($\lesssim 12$$R_\mathrm{S}$; $R_\mathrm{S}\equiv$ Saturn radius $ =6.0\times10^9\,\mathrm{cm}$), constrained by HST observations of the extended OH cloud (see their Figure 2, and references therein).  They solved the rate equations for number densities while also solving the radial diffusion equation, but energy conservation was not considered.  \cite{2002GeoRL..29x..25J} and \cite{jurac2005} further improved on these models by considering neutral cloud expansion, and solved for plasma and neutral distributions self-consistently in order to study the source of water within Saturn's inner magnetosphere.  

Our model concentrates on the molecular chemistry.  We start with a uniform box and characterize transport by just a time scale.  However, we do consider energy balance.  More importantly, unlike the above models, we retain H$_3$O$^+$ in our model, which proves to be a significant component.

The purpose of this paper is to summarize the sensitivity of the chemistry of Enceladus's torus to several parameters.  The parameters we  investigated are hot-electron temperature, hot-electron density, H$_2$O source rate, ion radial-diffusion time scale, and proton temperature anisotropy.  Being lighter, protons are less bound to the equator \citep{1980GeoRL...7...41B,2008P&SS...56....3S,2009JGRA..11404211P}.  This means protons spend only a fraction of the time interacting with the heavy ions and molecules.  This effect is simulated with a `proton dilution' factor (Section \ref{modl}).  We search for values of these parameters leading to thermal-electron temperature, thermal-electron density, and water-group ($\mathrm{W^ +\equiv O^++OH^++H_2O^++H_3O^+}$) ion-to-proton ratio consistent with available Cassini data (Section \ref{constraints}).

The observations used to constrain our model and to define the parameter space are given in Section \ref{data}.  The model is described in Section \ref{modl}.  The best fit (baseline solution) and the procedure used to find it are discussed in Section \ref{BaselineSolution}.  Model sensitivity to each of the parameters is discussed in Section \ref{grdSrch}.  Finally, the importance of hot electrons with regard to water-group ion composition is demonstrated in Section \ref{fehModulation}.
\section{Observations}\label{data}
\subsection{Parameters}\label{parameterData}
Here we present the observations used to bound the parameter search (Section \ref{modl}).  The baseline parameter values (listed in Table \ref{nominalOutput}) are mentioned throughout this section and are discussed at length in Section \ref{BaselineSolution}.

\paragraph{Neutral Source ($N_\mathrm{src}$)} We have set the baseline source of water to be $N_\mathrm{scr}=2.0\times10^{-4}$ molecules \,cm$^{-3}$\,s$^{-1}$.  A torus centered on Enceladus with cross section $(2R_\mathrm{S})^2$ has a volume of $\approx2\pi(4R_\mathrm{S})(2R_\mathrm{S})^2=2.2\times10^{31}$ cm$^3$, giving a volumetric source rate of $4.4\times10^{27}$ H$_2$O molecules s$^{-1}$, or  130 kg s$^{-1}$.  If one chooses a smaller or larger torus volume, the net neutral source rate is adjusted accordingly.  The best estimates of the Enceladus source come from Cassini observations.  \cite{2006Sci...311.1422H} estimated $150\lesssim N_\mathrm{src}/(\mathrm{kg\ s^{-1}})\lesssim 350$ from two stellar occultation observations of the Enceladus plume with Cassini UVIS.  \cite{2006Sci...311.1409T} inferred a source rate of 100 kg s$^{-1}$ from the CAssini Plasma Spectrometer (CAPS) data acquired during the E2 Cassini flyby of Enceladus on 14 July 2005.

Earlier source estimates come from models of the neutral clouds.  \cite{2002GeoRL..29x..25J} included the effects of collisional heating and developed a model to simulate the morphology of the extended OH cloud.  They used neutral lifetimes derived from a two-dimensional model by \cite{1998JGR...10320245R} to determine the neutral H$_2$O source responsible for the OH radial profile constrained by 1996 HST faint-object spectrograph observations.   They found a total water source required to maintain the OH cloud of $N_\mathrm{src}=112$ kg s$^{-1}$, 93 kg s$^{-1}$ coming from the orbit of Enceladus. 

In a later paper, \cite{jurac2005} improved their model, treating plasma and neutrals self-consistently by tracking neutrals with a Monte Carlo algorithm and transporting plasma diffusively.  They found a total water source rate of $\approx 300$ kg s$^{-1}$.  A similar result was found by \cite{burger2007} with a three-dimensional Monte Carlo simulation of neutrals constructed to simultaneously model the Ion and Neutral Mass Spectrometer (INMS) and UVIS observations made during the E2 Cassini flyby of Enceladus.

\paragraph{Radial Transport Time Scale ($\tau_\mathrm{trans}$)} We have set the baseline radial transport time scale to be $\tau^\mathrm{baseline}_\mathrm{trans}$\,=\,26\,days.  \cite{1998JGR...10320245R} estimated $\tau_\mathrm{trans}\approx$ 23 days from Voyager-era data and HST OH observations.  Using the radial velocities of \cite{2006P&SS...54.1197S} (their Figure 9), one finds a diffusion time scale at Enceladus of $\approx 2R_\mathrm{S}/v_\mathrm{R}(r=4R_\mathrm{S}) \approx$ 12 days.  This more rapid diffusion might suggest that the magnetosphere was compressed on Saturn Orbital Insertion (SOI), which would increase the angular speed to beyond corotation as momentum is conserved, ultimately resulting in enhanced radial outflow velocities (reduced transport times).  Thus, the \cite{2006P&SS...54.1197S} radial velocities based on SOI data may not represent the entire magnetosphere, and not for all epochs.  Radial convection may also be superimposed on the diffusive motions throughout Saturn's inner magnetosphere due to flux tube interchange instabilities \citep{2008JGRA..11301201R}.  A major goal of this study is to explore the consequences of such a wide range of time scales for radial transport.

\paragraph{Hot-electron Temperature and Fraction ($T_\mathrm{eh}$, $f_\mathrm{eh}$)}  Our model gives a baseline value for the temperature of hot electrons of $T^\mathrm{baseline}_\mathrm{eh}=160$ eV.  The Cassini Radio and Plasma Wave Science (RPWS) measurements by \cite{2005GeoRL..3220S02M} indicated a hot-electron component within 3--5$R_\mathrm{S}$ with $40\lesssim T_\mathrm{eh}/(\mathrm{eV})\lesssim90$.  From Cassini CAPS ELectron Spectrometer (ELS) observations acquired during the SOI period, \cite{2005Sci...307.1262Y} found that $T_\mathrm{eh}$ ranged from $\approx 500$--1000 eV inbound and $\approx 800$--3000 eV outbound (SOI), indicating a strong longitudinal and/or temporal dependence.  \cite{2008P&SS...56..901L} performed moment calculations on CAPS--ELS data with two different methods and found $300\lesssim T_\mathrm{eh}/(\mathrm{eV})\lesssim2000$ eV and $1500\lesssim T_\mathrm{eh}/(\mathrm{eV})\lesssim4000$.  \cite{Schippers2008} combined CAPS--ELS and Cassini Magnetospheric IMaging Instruement (MIMI) data and found significant scatter in $T_\mathrm{eh}$ and calculated $200\lesssim T_\mathrm{eh}/(\mathrm{eV})\lesssim 2000$ at the closest approach of $5.4R_\mathrm{S}$.  
 
Our baseline value for the fraction of the electron density in the hot component is found to be $f^\mathrm{baseline}_\mathrm{eh}=$ 0.46\%.  \cite{2005Sci...307.1262Y} found that $f_\mathrm{eh}$ ranged between 0.01\% and 5\% within 3--5$R_\mathrm{S}$.  As with $T_\mathrm{eh}$, $f_\mathrm{eh}$ varied significantly between the inbound and the outbound data.  \cite{2008P&SS...56..901L} found $f_\mathrm{eh}$ from their 3-D moment calculation to be $\lesssim$ 1\%.  \cite{Schippers2008} calculated $f_\mathrm{eh}$ as low as 0.1\% (inbound) and as high as 0.3\% (outbound).

At the time of \cite{2008P&SS...56....3S}, the authors felt that hot-electron parameters measured by CAPS ELS were highly uncertain in the vicinity of Enceladus, due to penetrating radiation.  Instead, they used the \cite{2005GeoRL..3220S02M} RPWS observations, which found that $T_\mathrm{eh} \approx$ 50 eV.  For the hot-electron density, they used $n_\mathrm{eh} \approx 0.1\ \mathrm{cm}^{-3}$ from the \cite{sittler1983} Voyager observations (which are not affected by penetrating radiation).  \cite{2008P&SS...56....3S} combined the \cite{2005GeoRL..3220S02M} Cassini RPWS data for $T_\mathrm{eh}$ and the Voyager data for $n_\mathrm{eh}$ to compute the total (effective) electron temperatures $T_\mathrm{e}$.  Since Voyager and Cassini SOI were so different in time, they used only the total electron temperatures for their reaction rates.

\paragraph{Proton Dilution ($f_\mathrm{H^+}$)}
Previous work by \cite{2008P&SS...56....3S} (their Figure 4) showed that $\approx$\,2/3 of the proton population are distributed within a distance of $\approx1R_\mathrm{S}$ from the centrifugal equator at the orbit of Enceladus.  Protons are pulled above the equator by the ambipolar electric field and do not couple efficiently to the heavy water-group ions.  

In Section \ref{varyingProtonDilution}, we show that the proton abundance is strongly coupled to the hot-electron population via impact ionization.  Our model is consistent with any value of the proton dilution parameter $f_\mathrm{H^+}$ between 0.7 and 1.0.  Values of $f_{\mathrm{H}^+}=1$ or 0 represents the cases where no or all newly created protons are excluded from the model.  Thus, to simplify the present analysis, $f_{\mathrm{H}^+}^\mathrm{baseline}$ has been set to unity.


\subsection{Constraints ($n_\mathrm{e}$, $T_\mathrm{e}$, W$^+$/H$^+$)}\label{constraints}
Parameter combinations are evaluated by comparing the corresponding model output to the following observations.  We choose $n_\mathrm{e}=60$\,cm$^{-3}$, $T_\mathrm{e}=2$\,eV, and W$^+$/H$^+ = 12$ as the initial model constraints (Section \ref{BaselineSolution}).

\paragraph{Total Electron Density ($n_\mathrm{e}$)}
\cite{2005Sci...307.1255G} reported $20\lesssim n_\mathrm{e}/(\mathrm{cm}^{-3})\lesssim100$ within 3--5$R_\mathrm{S}$ during the approach and first orbit around Saturn (SOI) from the Upper Hybrid resonance Frequency (UHF), acquired by the RPWS instrument.  \cite{2005GeoRL..3220S02M} considered RPWS Quasi-Thermal Noise (QTN) on SOI and found $40\lesssim n_\mathrm{e}/(\mathrm{cm}^{-3})\lesssim70$ within 3--5$R_\mathrm{S}$.  \cite{2005GeoRL..3223105P} used  the RPWS UHF to determine $n_\mathrm{e}$ from five later orbits.  They discovered variability inside $\approx5R_\mathrm{S}$, with $n_\mathrm{e}$ ranging from 35--105 $\mathrm{cm}^{-3}$.  In a later paper, \cite{2009JGRA..11404211P} developed a diffusive equilibrium model from RPWS and CAPS data acquired on 50 passes through Saturn's inner magnetosphere from 30 June 2004 to 30 September 2007, and calculated an electron density of over 50\,cm$^{-3}$ near the orbit of Enceladus.

\cite{2008P&SS...56..901L} derived $6\lesssim n_\mathrm{e}/(\mathrm{cm}^{-3})\lesssim20$ within 3--5$R_\mathrm{S}$ from the moment calculations with CAPS--ELS electron distribution data.  Their analysis was based on SOI data when CAPS was not fully actuating \citep{2006P&SS...54.1197S,sittler2007}.  Also, Cassini is a three-axis stabilized spacecraft with a fixed field of view.  Hence, sampling is limited, and moment calculations are not as straight forward as they are for spinning spacecraft, as discussed by \cite{2008P&SS...56..901L}.  Additionally, Cassini was likely negatively charged on SOI \citep{2005Sci...307.1262Y}, resulting in a lower-than-expected $n_\mathrm{e}$ from the moment calculations.

\cite{Schippers2008} performed a multi-instrument analysis of the electron populations for several orbits and found $n_\mathrm{e}\approx 10$ cm$^{-3}$ from their own CAPS--ELS analysis, but admitted that a negative spacecraft potential inside 9$R_\mathrm{S}$ likely resulted in an underestimate of the thermal-electron density.  Instead they used the RPWS UHF analysis by \cite{Gurnett2004} in this region, where $n_\mathrm{e} \approx50$ cm$^{-3}$ near the orbit of Enceladus.

From consideration of the above-cited observations, we take a value of $n_\mathrm{e}=60$\,cm$^{-3}$ for the total electron density in our model.

\paragraph{Thermal-electron Temperature ($T_\mathrm{e}$)}
\cite{2005GeoRL..3220S02M} derived $1\lesssim T_\mathrm{e}/(\mathrm{eV})\lesssim4$ within 3--5$R_\mathrm{S}$ from RPWS data.  The SOI CAPS analysis by~\cite{2005Sci...307.1262Y} found that $2\lesssim T_\mathrm{e}/(\mathrm{eV})\lesssim20$.  By comparison, the CAPS analysis by~\cite{2008P&SS...56..901L} found that $T_\mathrm{e}$ ranged from roughly 2 to 4 eV within 3--5$R_\mathrm{S}$ using one method, and between 1 and 4 eV using another.  \cite{2006P&SS...54.1197S,sittler2007} found $T_\mathrm{e}\approx 1.5$ eV between 3.5 and 4.5 $R_\mathrm{S}$.  \cite{Schippers2008} calculated $T_\mathrm{e}\approx 2$ eV at closest approach ($5.4R_\mathrm{S}$) from a multi-instrument analysis.  This value of $T_\mathrm{e}= 2$ eV is the one we use for the thermal-electron temperature in our model.

\paragraph{Ion Composition (W$^+$/H$^+$)}\cite{2008P&SS...56....3S} calculated H$^+$ and W$^+$ densities from the CAPS SOI results of \cite{2006P&SS...54.1197S,sittler2007}.  Their Figures 5 and 6 show that W$^+$/H$^+$ ranged from $\approx 3$--15 over the torus (3--5$R_\mathrm{S}$).  The CAPS SOI analysis of \cite{2005Sci...307.1262Y} found $6\lesssim \mathrm{W^+/H}^+\lesssim 35$ within 3--5$R_\mathrm{S}$.  The \cite{2005Sci...307.1262Y} analysis was based on non-coincident CAPS IMS singles data, whereas the \cite{2008P&SS...56....3S} results were based on coincident time-of-flight data.  It is challenging to distinguish protons from water-group ions in the singles data, while protons are well-separated  from the water-group ions in their respective time-of-flight channels.  Observations based on IMS singles data are thus likely to overestimate the $\mathrm{W^+/H}^+$ ratio.

\cite{wilson2008} presented a forward modeling technique to CAPS data for dayside equatorial orbits between 5.5 and 11$R_\mathrm{S}$ to calculate ion densities and temperatures.   Extrapolating their results down to 5$R_\mathrm{S}$ gives W$^+$/H$^+\approx 15$.  \cite{2009JGRA..11404211P} used anisotropy measurements from CAPS and temperature measurements from RPWS and CAPS to estimate equilibrium distributions within Saturn's inner magnetosphere.  Near the Enceladus torus, they found W$^+$/H$^+$\,$\approx$\,10. We choose a value of W$^+$/H$^+$ = 12 for the water-group to proton density ratio in our model.


\section{Model}\label{modl}
Here we present a model based on the Neutral Cloud Theory (NCT) model described in \cite{2003JGRA..108.1276D}, which was developed to address the variability of plasma conditions in the Io torus.  The Delamere model was based on the earlier NCT models of \cite{1988JGR....93.1773S}, \cite{1994ApJ...430..376B}, \cite{1998JGR...10319901S}, and \cite{2001JGR...10629899L}.  The tools developed by \cite{2003JGRA..108.1276D} to study sensitivity of the plasma-dominated environment at Io are utilized here to study the neutral-dominated water-based Enceladus torus.  

The model is 0-dimensional and homogeneous.  In this paper, we calculate steady-state densities and temperatures of ions and neutrals originating from a pure H$_2$O source.  When volumetric quantities are reported, we have adopted the volume used by \cite{2007GeoRL..3409105D} [$2\pi(4R_\mathrm{S})(2R_\mathrm{S})^2\approx2\times 10^{31}$\,cm$^3$], which roughly corresponds to a torus of minor radius $1R_\mathrm{S}$ centered on Enceladus's orbit at $4R_\mathrm{S}$.  Because pickup temperatures vary with radial distance along Saturn's equatorial plane (appendix, Eqs.\ 6 and 7), the scaling is only approximately valid for the span considered here of 3--5$R_\mathrm{S}$.

The basic equations \citep{1983ApJ...274..429B} for number density and energy density for species $\alpha$ are
\begin{equation}
\frac{\partial n_\alpha}{\partial t}=\mathcal{S}_{\mathrm{m},\alpha}-\mathcal{L}_{\mathrm{m},\alpha}
\label{densEq}
\end{equation} 
and
\begin{equation}
\frac{\partial(\frac{3}{2} n_\alpha T_\alpha)}{\partial t}=\mathcal{S}_{\mathrm{E},\alpha}-\mathcal{L}_{\mathrm{E},\alpha.}
\label{energyDensEq}
\end{equation} 
The $\mathcal{S}_\alpha$'s and $\mathcal{L}_\alpha$'s represent source rates and loss rates, respectively, for species $\alpha$.  Following the convention of \cite{2003JGRA..108.1276D}, the factor of 3/2 in Eq.\ \ref{energyDensEq} will be dropped henceforth, so that we are actually solving for an `effective' temperature rather than energy.  The temperature is described as effective because for pickup ions, the temperatures perpendicular to the magnetic field are expected to be greater than the parallel temperature (for observations and further discussion see \cite{richardson1990,2005GeoRL..3220S02M,2006P&SS...54.1197S,sittler2007,tokar2008,2009JGRA..11404211P}).  A complete discussion of Eqs.\ \ref{densEq} and \ref{energyDensEq} for ions, electrons, and neutrals can be found in the appendix.

The rate at which the H$_2$O particles are introduced into the model ($N_\mathrm{src}$) is a free parameter.  Chemical pathways encompass a set of reactions involving H, H$^+$, H$_2$, H$_2^+$, O, O$^+$, O$^{++}$, OH, OH$^+$, H$_2$O, H$_2$O$^+$, and H$_3$O$^+$ (see the appendix for a complete list of reactions).  Note that we assume 50\% of the hydrogen produced from impact dissociation of H$_2$O and OH has enough energy to escape our model \citep{1998JGR...10320245R}.  Neutrals are assumed to be cold, having only bulk motion.  They are not collisionally-heated in the model, and here we assume that neutrals created from ion charge exchange have velocities greater than the escape speed from Saturn and are ejected from the model.  Steady-state number and energy densities are found for each species by solving Eqs.\ \ref{densEq} and \ref{energyDensEq} iteratively using a modified Euler method with second-order accuracy.  

At the heart of this work is a sensitivity investigation of model output within the parameter space spanned by:
\begin{tabbing}
\noindent $-$ \ \ Neutral source rate ($N_\mathrm{src}$/10$^{-4}$\,cm$^{-3}$\,s$^{-1}$): \ \= 0.2 \ \ \ \= $\rightarrow$ 3.0 \\
\noindent  $-$ \ \ Hot-electron temperature ($T_\mathrm{eh}$/eV): \> 20 \> $\rightarrow$ 400\\
\noindent  $-$ \ \ Hot-electron fraction ($f_\mathrm{eh}\equiv n_\mathrm{eh}/n_\mathrm{e}$): \> 0.05 \> $\rightarrow$ 1.0\% \\
\noindent  $-$ \ \ Radial transport time scale ($\tau_\mathrm{trans}$/days): \> 2 \> $\rightarrow$ 60\\
\noindent  $-$ \ \ Proton dilution factor ($f_{\mathrm{H}^+}$): \> 0.7 \> $\rightarrow$ 1.0.
\end{tabbing}
\noindent These ranges reflect the broad set of observations given in Section \ref{parameterData}.  The proton dilution factor, $f_{\mathrm{H}^+}$, removes protons from the model, and has been implemented by modifying the source term in Eq.\ \ref{densEq} for protons:
\begin{equation}
\frac{\partial n_\mathrm{H^+}}{\partial t}=f_{\mathrm{H}^+}\mathcal{S}_{\mathrm{m},\mathrm{H^+}}-\mathcal{L}_{\mathrm{m},\mathrm{H^+}}.
\label{protonDil}
\end{equation} 
\noindent In reality, the heavy ion abundance peaks near the centrifugal equator, while the proton abundance peaks well away from the equator and out of our model domain \citep{2009JGRA..11404211P}.  We apply the above equation to crudely address and investigate this phenomenon.  
 

\section{Baseline Solution}\label{BaselineSolution}
\subsection{Procedure}\label{proc}
Initially we set $f_\mathrm{H^+}=1$ (see further discussion in Section \ref{varyingProtonDilution}) and explore the space spanned by $f_\mathrm{eh}$, $T_\mathrm{eh}$, $\tau_\mathrm{trans}$, and $N_\mathrm{src}$ alone.  The model was run with a random-walk Monte Carlo algorithm to find the parameter space coordinates yielding the best agreement between model output and the constraints.  The following have been chosen as the constraints on the model (Section \ref{constraints}):  $n_\mathrm{e}=60$\,cm$^{-3}$, $T_\mathrm{e}=2$\,eV, $\mathrm{W^+/H^+=12}$.  In Section \ref{alternativeConstraints} we accommodate a wider range of observations and investigate how composition is affected by these choices of $n_\mathrm{e}$, $T_\mathrm{e}$, $\mathrm{W^+/H^+}$.  

We define best agreement as the smallest total fractional difference between the model output and the constraints: 
\begin{equation}
f_\mathrm{diff}=\sum_i\Bigg |1-\frac{\mathrm{Model}_i}{\mathrm{Constraint}_i}\Bigg |_,
\label{fracDiff}
\end{equation} 
where $i=n_\mathrm{e}$, $T_\mathrm{e}$, $\mathrm{W^+/H^+}$ in the present case.  The baseline solution (parameter combination with the smallest $f_\mathrm{diff}$) was found by starting the random-walk algorithm from a point in parameter space near the global minimum in $f_\mathrm{diff}$.  The procedure used to find the global minimum is discussed in Section \ref{grdSrch}.  Model output was evaluated and a step was randomly taken in the direction of one of the four parameters.  The step sizes were also random in length and constraint-dependent.  For example, the step size for the hot-electron fraction ranged from 0--0.001, while the step size for the transport time ranged from 0--86,400 seconds (1 day).  Every value throughout the step intervals had equal weight.  The model output was then evaluated ($f_\mathrm{diff}$ calculated), and the procedure was repeated until a minimum in $f_\mathrm{diff}$ was found.  The random walk led to the baseline parameter combination, where $f^\mathrm{baseline}_\mathrm{diff}\approx0$.  

The solution and corresponding model output are given in Table~\ref{nominalOutput}.  Lists of the values of lifetimes of each species controlled by the primary source/losses mechanisms are presented in Table \ref{time scalesMech}, and lifetimes listed by each separate reaction can be found in the appendix.  

\subsection{Results}\label{baselineSolutionResults}
  Table~\ref{nominalOutput} presents the model output for the densities and temperatures of all species.  We find a torus composition that is dominated by neutral species with roughly equal amounts of H, O and OH ($\approx$\,700 cm$^{-3}$ each) with a lesser amount of water molecules (190 cm$^{-3}$) and trace amounts of H$_2$. The ion species are dominated by H$_2$O$^+$ and OH$^+$ ($\approx$\,20 cm$^{-3}$) followed by H$_3$O$^+$ and O$^+$ ($\approx$\,10 cm$^{-3}$), H$^+$ ($\approx$\,5 cm$^{-3}$) and trace amounts of O$^{++}$ and H$_2^+$. 
 
When we compare our baseline OH density (770\,cm$^{-3}$) to Figure 3 of \cite{2002GeoRL..29x..25J} we find very similar values.  Their model was developed to simulate the morphology of Saturn's extended OH cloud, as measured by HST, October 2002.  They found an OH density of $\gtrsim$\,750\,cm$^{-3}$ centered on the orbit of Enceladus (see also \cite{jurac2005}).  This is an independent test of our results since no radiative constraints were used to determine the baseline solution.

Our model shows that all ion species have temperatures close to their initial pickup temperature, consistent with negligible loss of energy via radiation and Coulomb collisions with electrons.  The thermal coupling time between electrons and ions derived from the model is $\approx$\,60 days.  Because ions are transported out of the box in 26 days, they do not efficiently transfer energy to the electrons.  Oxygen ions are picked up by the corotating magnetosphere with a temperature of 38.4\,eV (appendix, Eq.\ 7).  OH, H$_2$O, and H$_3$O ions are picked up with 40.8, 43.2, and 45.6 eV, respectively.  As shown in Table \ref{nominalOutput}, very little of these heavy ions' thermal energy has been transferred to the thermal electrons.  This result has also been established in Figure \ref{enerFlow} by the small energy coupling (2.4\%) between ions and electrons.  

The water-group temperatures from our model ($\approx$\,38--42 eV) are warmer than the CAPS data suggest.  \cite{2006P&SS...54.1197S,sittler2007} observed that $T_{\perp,\mathrm{W^+}}$\,$\approx$\,35--40 eV near the Encaladus torus.  With their anisotropy of $(T_\perp/T_\parallel)_\mathrm{W^+}$\,$\approx$\,5, the effective water-group temperature is reduced to $T_\mathrm{W^+}=(2T_\perp+T_\parallel)/3 \approx$\,27 eV.  The discrepancy between our model and the data may be explained by a sub-corotating plasma torus near Enceladus' $L$ shell.  An $\approx$\,20\% sub-corotation of the plasma flow, as measured by CAPS at 4$R_\mathrm{S}$ (R. Wilson, personal comm.), would reduce pickup energies and may account for the difference between our model temperatures and the \cite{2006P&SS...54.1197S,sittler2007} observations (40 eV and 27 eV, respectively). 

In our model, we have assumed that the ion velocity distributions are isotropic and thus cannot comment on the parallel and perpendicular temperatures individually (Section \ref{modl}).  Because the data suggest that the water-group ion has an anisotropy of $T_\perp/T_\parallel \approx 5$ and the protons have an anisotropy of $T_\perp/T_\parallel \approx 2$ \citep{richardson1990,2005GeoRL..3220S02M,2006P&SS...54.1197S,sittler2007}, we hope to include anisotropic ion velocity distributions in the future.

The flow of mass and energy is shown in Figures \ref{partFlow} and \ref{enerFlow} (contributions from individual species are listed in the appendix).  Mass is introduced into the model by way of H$_2$O only and leaves when ions radially diffuse, charge exchange with neutrals, or recombine with electrons (for singly-ionized species).  Recombination and charge exchange represent mass sinks because the ions become `fast neutrals,' assumed to possess enough velocity to escape the torus.  We find that 94\% of the particles leave the torus as fast neutrals and 6\% as diffusing ions.

Energy is introduced almost entirely by pickup ions, and a small amount (1.8\%) comes from Coulomb collisions between the hot electrons and the thermal electrons/ion species.  Pickup ions represent an energy source due to the velocity difference between neutrals and ions in the Enceladus torus; a freshly-ionized neutral is accelerated to magnetospheric corotation via current systems established between Saturn's ionosphere and the Enceladus torus.  Fresh pickup ions may be produced either by electron impact or charge exchange between an ion and a neutral.  Energy is carried away in the model by the fast neutrals, diffused ions, and radiation induced by electron-impact excitation.  Figure \ref{enerFlow} indicates that most energy (83\%) leaves the model with the fast neutrals.

Figures \ref{partFlow} and \ref{enerFlow} can be compared to the energy and particle flow diagrams produced by the oxygen-based model of \cite{2007GeoRL..3409105D} (their Figure 1).  In the partitioning of particle flow between radial transport and fast neutrals, their simplified model is remarkably similar to our solution, though their neutral source rate (4\,$\times$\,10$^{-4}$\,cm$^{-3}$\,s$^{-1}$) is twice as strong.

The total energy flowing through our model is roughly 40\% of that found by \cite{2007GeoRL..3409105D} (9.4\,eV\,cm$^{-3}$\,s$^{-1}$ compared to 23\,eV\,cm$^{-3}$\,s$^{-1}$).  They used a smaller hot-electron fraction ($f_\mathrm{eh}=0.3$\%) and a higher hot-electron temperature ($T_\mathrm{eh}=1000$\,eV) than the values we used to produce Figures \ref{partFlow} and \ref{enerFlow}.  In addition, their transport time was considerably longer at 45 days.

Regarding energy output, we find that more energy is transported out of the torus by ions than by fast neutrals when compared to \cite{2007GeoRL..3409105D}, though we agree on the radiated power partition of a few percent.  Energy input is remarkably different because we have included molecular chemistry, while \cite{2007GeoRL..3409105D} included atomic oxygen only.   Hot-electron thermal coupling plays a bigger role as an energy source with 1.8\% of the total energy input compared to 0.5\% in \cite{2007GeoRL..3409105D}.  In our model, charge exchange only marginally exceeds the combination of photoionization and impact ionization as a means of adding fresh pickup ions to the system, whereas in \cite{2007GeoRL..3409105D} charge exchange was found to dominate these mechanisms by a factor of 19.  One reason for this major difference is the relative ease at which OH is ionized via electron impact with respect to O.  This reaction can contribute greatly to the overall energy budget since OH is the dominant species (Table \ref{nominalOutput}).  Photoionization of OH and O occurs at roughly the same, much lower rate for the baseline case.  If oxygen were the only species in the model, energy input would be a competition between the highly-likely resonant charge exchange between O and O$^+$ and the order-of-magnitude-less-likely photo- plus impact ionization of O.  The reaction rates supporting the above argument can be found in the appendix.

Dissociative recombination represents an important plasma sink to the Enceladus torus.  This process has a profound effect on the neutral-to-ion ratio in the torus.  The \cite{2007GeoRL..3409105D} model found $n_\mathrm{neut}/n_\mathrm{ions}=12$, but they argued that this ratio represents a lower limit since their model is oxygen-based and does not include dissociative-recombination reactions; the recombining ions' neutral products have escape velocities and leave the model, just as recombining atomic ion species do.  We find $n_\mathrm{neut}/n_\mathrm{ions}=40$ with the full water-based molecular chemistry.  

\cite{2008P&SS...56....3S} showed that the combination of cold electrons ($\approx$ 1 eV) and the dominance of molecular ions over atomic oxygen ions near Enceladus \citep{2005Sci...307.1262Y} drives a rapid dissociative-recombination time scale.  They also showed that this process increases the neutral-to-ion production ratio to $\approx$\,50 near Enceladus (their Figures 18 and 20).  

To illustrate the importance of dissociative recombination in our model, we increased $T_\mathrm{e}$ by increasing the pickup temperature in the Enceladus torus (appendix, Eq.\ 7).  Indeed, if the parameters are fixed at the baseline values (Table \ref{nominalOutput}), and $T_\mathrm{e}$ is increased to 6 eV (as in the ion-dominated Io torus), dissociative recombination continues to prevent ions from dominating neutrals (Section \ref{dominantChemistry}).  At Saturn, impact ionization by thermal electrons cannot compete with dissociative recombination as an ion sink, even when $T_\mathrm{e}=$ 6 eV.

  \paragraph{Derived Quantities}  The field strength in the Enceladus torus is 325\,nT \citep{2006LPI....37.1585D}, giving a plasma beta of
\begin{equation}
\beta= \frac{\sum_{j=\mathrm{ions,e,eh}}{n_jT_j}}{B^2/8\pi}=0.0091\approx1\%,
\label{plasmaBeta}
\end{equation} 
where the summation is taken over all charged species, including both the thermal- and hot-electron populations (Table \ref{nominalOutput}).  This plasma beta is consistent with the \cite{2008P&SS...56....3S} analysis, which found $\beta$ between 3 and 5$R_\mathrm{S}$ to be between 0.1 and 5\%.  The Alfv\'{e}n speed is given by
\begin{equation}
v_\mathrm{A}=B/\sqrt{4\pi\rho}=230\,\mathrm{km}\,\mathrm{s}^{-1},
\label{alfvenSpeed}
\end{equation} 
where $\rho=\sum_{j=\mathrm{ions}}n_jm_j$.  If the plasma is at full corotation at the orbit of Enceladus [as has been assumed in the model for calculating $E_\mathrm{pu}$ (appendix, Eq.\ 7)], then the Alfv\'{e}n Mach number is 
\begin{equation}
M_\mathrm{A}=v_\phi/v_\mathrm{A}=(4R_\mathrm{S})\Omega_\mathrm{S}/v_\mathrm{A}=0.17.
\label{alfvenMach}
\end{equation} 
This should be compared to~\cite{2008P&SS...56....3S}, who find $0.01<M_\mathrm{A}<0.5$ between 3 and 5$R_\mathrm{S}$.  \cite{2008P&SS...56....3S} use ion--electron fluid parameters as boundary conditions to solve for ion field-line distributions throughout Saturn's inner-magnetosphere.  The fluid parameters are derived from CAPS data acquired during the approach phase of the SOI period~\citep{2006P&SS...54.1197S,sittler2007}.  In a future study, we will solve the radial-transport equation and present a self-consistent map of the ion distribution throughout Saturn's inner magnetosphere (as performed by \cite{1998JGR...10320245R}).  

\subsubsection{Lifetimes}\label{losstime scales}
The lifetimes are listed by mechanism in Table \ref{time scalesMech}.  The lifetimes are listed by reaction in the appendix.  The diffusion lifetime ($\tau_\mathrm{trans}$) has not been well-constrained by this study.  In Section \ref{grdSrch}, we find that the model is consistent with a diffusion time scale of 12 days and longer.  Because our study cannot place an upper limit on $\tau_\mathrm{trans}$, it is not possible to say which mechanisms occur more rapidly than radial transport and are therefore more important.

Our model does not account for collisional heating mechanisms that would give neutrals enough speed to escape the model domain.  Thus, we have assumed that the time scales for such neutral escape are longer than the time scales for the included mechanisms.  Based on the Enceladus torus study by \cite{farmer2008}, we now examine this assumption and conclude that the effects of collisional heating should be included in future iterations of our model, especially in the case of OH.  

Neutral H$_2$O has an effective cross-section to neutral--neutral collisions due to dipole--dipole interactions such that an H$_2$O molecule will be transported outside our modeled torus ($>$\,5$R_\mathrm{S}$) after $\approx$\,40\,days (Figure 2 in \cite{farmer2008}). Then, according to Table \ref{time scalesMech}, only impact dissociation and charge exchange are important loss mechanisms for H$_2$O.  A similar time scale (40 days) may also limit OH lifetimes because its induced dipole is comparable to that of H$_2$O.  Under this assumption, collisional heating would be the $most$ important loss mechanism for OH.

The \cite{farmer2008} result can also be used to estimate oxygen lifetimes against neutral--neutral collisions.  Because the geometric cross section for H$_2$O is a factor of 10 smaller than the induced-dipole cross section ($\sigma_\mathrm{H_2O}= 5$\,\AA$^2$\,$\rightarrow$\,$\sigma^\mathrm{ind}_\mathrm{H_2O}= 54$\,\AA$^2$, \cite{farmer2008}), and because the collision frequency is proportional to $\sigma$, one might expect that O will take roughly 10 times $longer$ to be scattered outside the torus.  That is, only those mechanisms occurring on a time scale of $\lesssim$\,400\,days (charge exchange and impact ionization) would occur before neutral collisions remove oxygen from the torus.

\subsubsection{Dominant Chemistry}\label{dominantChemistry}
Reactions occurring more frequently than $10^{-6}\,\mathrm{cm}^{-3}\,\mathrm{s}^{-1}$ , shown in Table \ref{reactionSummary}, are the ones of primary importance to the torus chemistry.  The appendix lists the full set of reactions and reaction rates for the baseline case.  For example, if we run the model with every reaction in the appendix turned on (and the parameters set at the baseline values) all densities  and ion temperatures are within 3\% of the results from the calculation using the refined set presented in Table \ref{reactionSummary}.

The most dominant reaction is impact dissociation of H$_2$O by hot electrons ($\mathrm{H_2O+e_{h}\rightarrow OH+H+e}$).  Dissociation of H$_2$O by hot electrons occurs so frequently because of the relatively large reaction rate at $T^\mathrm{baseline}_\mathrm{eh}=160$ eV (appendix) as well as the high baseline density of H$_2$O (Table \ref{nominalOutput}).  Several other charge-exchange, photolytic, and electron-impact reactions are competitive behind H$_2$O impact dissociation.  

Impact ionization by hot electrons contributes roughly the same amount of energy (22\% of total input) via magnetospheric pickup as does photoionization (16\%). Impact ionization by thermal electrons is a minor source of energy to the torus ($\approx 1\%$).  In fact, only one reaction in Table \ref{reactionSummary} involves thermal-electron impact ionization ($\mathrm{OH+e\rightarrow OH^+ +2e}$).  \cite{2007GeoRL..3409105D} find that charge exchange is far more important than photo- and impact ionization combined as a torus energy source.  Here we find that charge exchange is only marginally more important ($\approx 60\%$) than the combination of these other ionization sources ($\approx 40\%$) for providing fresh pickup ions to the torus.  This discrepancy (discussed in Section \ref{baselineSolutionResults}) is largely due to the fact that the earlier model did not include molecular chemistry.

  \subsection{Sensitivity}\label{baselineSolutionSensitivity}
Contour plots of the total fractional difference ($f_\mathrm{diff}$, Eq.\ \ref{fracDiff}) between the model output and the constraints have been created for every parameter combination and are shown in Figure \ref{sensPlot}.  The intersection of the dashed lines indicates the baseline solution (Table \ref{nominalOutput}).  In each case, the remaining three parameters are fixed at the baseline values.  The shading inside the $f_\mathrm{diff}=1$ contour is intended to guide the eye when comparing one panel to another.  Because the remaining two parameters are fixed in each panel, these plots show the sensitivity of $f_\mathrm{diff}$ to each parameter individually.
  
The source rate is inversely related to $\tau_\mathrm{trans}$ (Panel 1) and $f_\mathrm{eh}$ (Panel 2).  The trend between $N_\mathrm{src}$ and $\tau_\mathrm{trans}$ can be understood as a balance between source and sink; plasma taking longer to diffuse out of the model must be accompanied by a decrease in H$_2$O.  Similarly, an increase in hot electrons results in higher ionization.  A smaller source rate is required to maintain the torus composition consistent with the constraints on $n_\mathrm{e}$ and W$^+$/H$^+$.
    
The total fractional difference, $f_\mathrm{diff}$, strongly depends on $f_\mathrm{eh}$, with a strongly pronounced valley in all cases (Panels 2,\,4,\,5).  The hot-electron population is critical for ionizing H efficiently to obtain $\mathrm{W^+/H^+\approx12}$ (and hence, minimizing $f_\mathrm{diff}$).  The hot electrons are equally necessary for attaining a higher overall ionized composition, thereby increasing $n_\mathrm{e}$, and for heating the thermal electrons ($T_\mathrm{e}$) via Coulomb coupling.

A similar dependence exists for $T_\mathrm{eh}$, except that the strong dependence is at low $T_\mathrm{eh}$ only (Panels 3,\,5,\,6).  Beyond $T^\mathrm{baseline}_\mathrm{eh}$ (160 eV), $f_\mathrm{diff}$ is roughly independent of $T_\mathrm{eh}$.  The other three parameters dominate variation in $f_\mathrm{diff}$ when the hot-electron population is sufficiently energetic ($\gtrsim 160$ eV).  

\cite{2007GeoRL..3409105D} present sensitivity contour plots of the neutral-to-ion ratio and thermal-electron temperature from their oxygen-based model (their Figure 2).  We have generated similar contours (not shown) and find that the torus composition is neutral-dominated despite $T_\mathrm{e}$ approaching 6 eV, whereas \cite{2007GeoRL..3409105D} find that ions dominate when $T_\mathrm{e}$ is as low as 3 eV.  The key difference between their model and ours is that we have included a complete water-based, molecular set of reactions, and we have included the effects of dissociative recombination not present in the Delamere model.

Because we are varying only two parameters at a time in each panel of Figure \ref{sensPlot}, we are not necessarily finding the highest quality of fit (smallest $f_\mathrm{diff}$) throughout.  This concern is addressed in Section \ref{grdSrch}.
  
\section{Grid Search}\label{grdSrch}
In search of the best possible match to our three observable constraints, we have explored the full 5-dimensional parameter space ($f_\mathrm{eh}$, $T_\mathrm{eh}$, $f_\mathrm{H^+}$, $N_\mathrm{src}$, $\tau_\mathrm{trans}$) by solving Eqs.\ \ref{densEq} and \ref{energyDensEq} systematically for a large set of parameter combinations.  We divided the domain for each parameter (Section \ref{modl}) into 30 discrete, uniformly-spaced values.  We created a table of model output values of densities and temperatures corresponding to every parameter combination.  This table, or grid, is an extremely useful tool for comparing output from our model to any combination of observations (Section \ref{alternativeConstraints}).  The computationally-expensive procedure of creating the grid will only need to be repeated when additional or updated chemical reactions are introduced to the model.  The downside of the grid search is limited resolution, and to double the resolution would require $2^5$ times more computational time.  Fortunately, the calculation can be done in parallel, and wall-clock time can be reduced linearly with the number of computer processors.  

\subsection{Results}\label{gridSearchResults}
In the present case, we have used the grid to generate sensitivity contours for every parameter combination by searching for the smallest $f_\mathrm{diff}$ everywhere using the constraints in Section \ref{BaselineSolution}:  $n_\mathrm{e}=$ 60\,cm$^{-3}$, $T_\mathrm{e}=$ 2\,eV, W$^+$/H$^+ =$ 12.  The primary purpose of these contours is to establish the following parameter limits consistent with the primary constraints:
\begin{enumerate}
  \item[\ ] $1.5\lesssim N_\mathrm{src}/(10^{-4}\mathrm{\ cm^{-3}\ s^{-1}})\lesssim 2.7$ 
  \item[\ ] $12\ \mathrm{days} \lesssim \tau_\mathrm{trans}$
  \item[\ ] $0.3\lesssim f_\mathrm{eh}(\%)\lesssim0.9$
  \item[\ ] $100\lesssim T_\mathrm{eh}/\mathrm{eV}\lesssim250$
  \item[\ ] $ f_\mathrm{H^+}\le1$.
\end{enumerate}
\noindent  We will also demonstrate that the solution space found by searching the grid is smoothly varying and well-behaved.

In Figure \ref{tehVfeh}, we present the sensitivity contours between the parameters $f_\mathrm{eh}$ and $T_\mathrm{eh}$.  Panel 1 is the contour plot for the smallest obtainable total fractional difference within the domain defined in Section \ref{modl}.  Recall that $f_\mathrm{diff}$ is the sum of fractional differences between model output and primary model constraints for $n_\mathrm{e}$, $T_\mathrm{e}$, and W$^+$/H$^+$ (Section \ref{data}).  To illustrate the individual contributions to $f_\mathrm{diff}$, we show $T_\mathrm{e}$, $n_\mathrm{e}$, and W$^+$/H$^+$ in Panels 2, 3, and 4 respectively; the dotted contours define the primary constraints used to evaluate $f_\mathrm{diff}$.  The values of the parameters $\tau_\mathrm{trans}$, $N_\mathrm{src}$, and $f_\mathrm{H^+}$ corresponding to the model output for $n_\mathrm{e}$, $T_\mathrm{e}$, and W$^+$/H$^+$ are plotted in Panels 5--7.  We also present the contours for model output quantities:  water-group composition (Panels 8--11), total UV power (Panel 12), and mass radial transport rate (Panel 13).  

The $f_\mathrm{diff}=0.05$ contour in Panel 1 has been shaded in gray and over-plotted in Panels 5--13 to indicate a parameter subspace consistent with the observations.  This contour can be expanded or reduced to reflect observational uncertainty.  The random-walk algorithm used in Section \ref{BaselineSolution} to find the baseline solution was initialized with parameter coordinates near the global minimum inside this solution space.  The baseline solution has been indicated throughout Figure \ref{tehVfeh} by the intersection of dashed lines.  Though the baseline solution is encompassed in the solution space, it is not unique.  The solution space is bounded by a range of the parameters for which a solution can be found that agrees with the data approximately as well as the baseline solution discussed in Section \ref{BaselineSolution}.  From Figure \ref{tehVfeh}, we find that $0.3\lesssim f_\mathrm{eh}(\%)\lesssim 0.9$ and $100\lesssim T_\mathrm{eh}/(\mathrm{eV})\lesssim 250$. 

Because parameter combinations are evaluated according to model output for $n_\mathrm{e}$, $T_\mathrm{e}$, and W$^+$/H$^+$ alone, one cannot rule out a priori that adjacent points in these contours sample wildly different values of $\tau_\mathrm{trans}$, $N_\mathrm{src}$, and $f_\mathrm{H^+}$.  That the contours for these constraints (Panels 5--7) and for the composition (Panels 8--13) are smoothly varying and well-behaved proves that this is not the case.

Because the different water-group ion species have similar masses and because composition varies significantly with longitude in the Enceladus region \citep{2008AGUFM.P23B1384W}, the water-group density ratios (Panels 8--11) are difficult to distinguish in the CAPS observations.  However, \cite{2008P&SS...56....3S} have shown, based on CAPS SOI data, that H$_3$O$^+$/W$^+$ $\approx 0.45$, H$_2$O$^+$/W$^+$ $\approx 0.15$, and OH$^+$/W$^+$ $\approx$ O$^+$/W$^+$ $\approx 0.2$ near the orbit of Enceladus.  Preliminary CAPS results by \cite{2008AGUFM.P23B1384W} also suggest that H$_3$O$^+$ is the dominant water-group species in the Enceladus torus.  

INMS data acquired downstream of Enceladus suggest that the local chemistry, dominated by charge exchange, may represent a significant source of H$_3$O$^+$ not included in our model \citep{cravens2009}.  We are currently preparing a manuscript on the chemical interaction between Enceladus's water-based plumes and the corotating plasma in which we identify efficient chemical pathways by which H$_3$O$^+$ is created.  We hope to incorporate this correction into our model and to properly compare our results to the CAPS data in a future study.

The UV power due to impact excitation (Panel 12) was calculated from line emissions and scales to roughly 1 GW for the entire torus.  This estimate likely represents a lower limit since emission from H$_2$O, for example, has not been considered in our model (see appendix).  Most of the radiated power ($\approx 80$\%) comes from the 1304, 1356, and 6300\,\AA\ neutral oxygen emission lines.  If measured, $P_\mathrm{UV}$ would be a powerful constraint for our model \citep{esposito2005}.

Radial transport, or mass loading ($\dot{M}$, Panel 13), is defined here as the total radially-transported mass:
\begin{equation}
\label{radialTransportRate}
\mathrm{Radial\ Transport\ Rate\ }\equiv \frac{\mathrm{Vol}}{\tau_\mathrm{trans}}\sum_{j=\mathrm{ion,e}} m_j n_j,
\end{equation} 
where a torus volume of $2\pi(4R_\mathrm{S})(2R_\mathrm{S})^2$ has been used, and $m_j$ is the mass of ion species $j$.  We find that ion transport may vary by a factor of six or more ($\approx 8$--50\,kg\,s$^{-1}$) and still be consistent with the solution space presented here.

The solution space can be further constrained when better data for the parameters themselves [$N_\mathrm{src},\ f_\mathrm{eh},\ T_\mathrm{eh},\ f_{\mathrm{H}^+},\ \tau_\mathrm{trans}$ (Section \ref{parameterData})] become available.  Limits derived from such measurements could be used to define contour levels in Panels 5--7 that would limit the solution space in gray.  The same can also be said for torus ultraviolet emission ($P_\mathrm{UV}$).

The fractional difference contours in Panel 1 of Figure \ref{tehVfeh} are much broader than in Panel 5 of Figure \ref{sensPlot}.  The difference of course being that $\tau_\mathrm{trans}$, $N_\mathrm{src}$, and $f_\mathrm{H^+}$ are fixed in Figure \ref{sensPlot} while in Figure \ref{tehVfeh} they are free.  To help understand how the free parameters (Panels 5--7) have expanded the solution space, we offer the following interpretation.  For clarity$-$and because the proton abundance is strongly coupled to hot-electrons (Section \ref{varyingProtonDilution})$-$we restrict our discussion to transport time and source rate.  Focus on the upper-right quadrant of Panels 5, 6 and 7 in Figure \ref{tehVfeh}.  This is the region where the total fractional difference contours (Panel 1) have been broadened most markedly when compared to Figure \ref{sensPlot}.  The transport time increases slightly ($\tau_\mathrm{trans}\uparrow$) with increasing hot-electron temperature ($T_\mathrm{eh}\uparrow$) and decreases sharply ($\tau_\mathrm{trans}\downarrow$) as the hot-electron fraction increases ($f_\mathrm{eh}\uparrow$).  The neutral source rate is remarkably constant at a value near the baseline value of $2.0\times10^{-4}$\,cm$^{-3}$\,s$^{-1}$, varying by only as much as 30\% throughout the entire upper-right quadrant.  

The trends between $\tau_\mathrm{trans}$, $f_\mathrm{eh}$, and $T_\mathrm{eh}$ are in part driven by the thermal-electron density.  As the hot electrons increase in temperature beyond the baseline hot-electron temperature, 160 eV ($T_\mathrm{eh}\uparrow$), the impact ionization rate of both hydrogen and water-group molecules decreases (Figure \ref{WandHimpIonRates}).  To maintain the total ionization ($n_\mathrm{e}$), the ions must remain in the torus longer ($\tau_\mathrm{trans}\uparrow$).  Similarly, as the hot electrons increase in $number$ ($f_\mathrm{eh}\uparrow$), both protons and water-group ions build up and must transport out of the model more $rapidly$ ($\tau_\mathrm{trans}\downarrow$) to maintain $n_\mathrm{e}$.

In Figure \ref{tauVfeh}, we also include the same set of sensitivity contours, this time between $\tau_\mathrm{trans}$ and $f_\mathrm{eh}$.  The solution space is bounded by $0.3\lesssim f_\mathrm{eh}(\%)\lesssim0.9$, but $\tau_\mathrm{trans}$ is only bounded from below at 12 days.  There are three other sensitivity plots in the appendix with which to constrain $\tau_\mathrm{trans}$, but $\tau_\mathrm{trans}$ is never constrained from above because recombination and charge-exchange dominate the chemistry when $\tau_\mathrm{trans}\gtrsim$ 26 days.  That is, varying $\tau_\mathrm{trans}$ in this regime has no effect on the model output.  The importance of recombination has also been discussed in \cite{2008P&SS...56....3S}.  

Sensitivity plots similar to Figures \ref{tehVfeh} and \ref{tauVfeh} for the eight remaining parameter combinations can be found in the appendix.  The $f_\mathrm{diff}=0.05$ shaded contours in the appendix have been used to find the limits on the neutral source rate ($N_\mathrm{src}$) and the proton dilution factor ($f_\mathrm{H^+}$) presented above and in Section \ref{conclusion}.

\subsubsection{Proton Dilution}\label{varyingProtonDilution}
The grid search has revealed that $f_\mathrm{H^+}$ is a weak parameter in the model.  That the model output is weakly dependent on $f_\mathrm{H^+}$ is illustrated in Figures \ref{tehVfeh} and \ref{tauVfeh}.  Panel 7 in each figure shows that small changes in $T_\mathrm{eh}$ and $f_\mathrm{eh}$ can compensate for large changes in $f_\mathrm{H^+}$.  In fact, the full domain of $f_\mathrm{H^+}$ (0.7--1.0) fits entirely into the shaded solution subspace.  Table \ref{reactionSummary} (and the reaction rates in the appendix) indicates that the dominant reaction for creating protons is impact ionization of hydrogen by hot electrons:
\begin{equation}
\label{HImpactIonization}
\mathrm{H+ e_\mathrm{h}\rightarrow H^+ + 2e}.
\end{equation} 
This reaction strongly couples $f_\mathrm{H^+}$ to $f_\mathrm{eh}$, reducing the parameter space to four dimensions.  The reaction H$_2$O + e$_\mathrm{h}$ $\rightarrow$ H$^+$ + OH + 2e is only about 1/3 as effective as hot-electron impact ionization at producing fresh H$^+$ (appendix), while O$^+$ + H $\rightarrow$ O + H$^+$ is independent of hot electrons altogether.

Figure \ref{fhpVoutput} illustrates that both the fit and the composition depend weakly on the choice of $f_\mathrm{H^+}$.  The grid was searched to find the best combination of $T_\mathrm{eh}$, $f_\mathrm{eh}$, $N_\mathrm{src}$, and $\tau_\mathrm{trans}$ for three different values of $f_\mathrm{H^+}$: 0.7, 0.85, 1.0.  The combination yielding the best agreement is shown in the left panel$-$normalized to the baseline values in Table \ref{nominalOutput}$-$and the model output for water-group composition is shown in the right panel.
 
\subsubsection{Varying Primary Constraints ($n_\mathrm{e}$, $T_\mathrm{e}$, W$^+$/H$^+$)}\label{alternativeConstraints}
The parameter space becomes 8-dimensional if variation in the primary constraints ($n_\mathrm{e}$, $T_\mathrm{e}$, and W$^+$/H$^+$) is allowed.  In this case we are interested in how composition is affected by allowing these primary constraints to take on values reflecting a wide range of observations (Section \ref{constraints}).  Such an exercise illustrates how the choice of constraints affects the baseline solution.  As with the baseline solution already discussed, we set $f_\mathrm{H^+}=1$ in all cases since the results are only mildly sensitive to $f_\mathrm{H^+}$ (Section \ref{varyingProtonDilution}).

To this end, the grid was searched to find parameter combinations consistent with the following six constraint combinations:
\begin{tabbing}
 \ \ $n_\mathrm{e}=40$\,cm$^{-3}$\ \ \ \ \ \ \= $T_\mathrm{e}=\mathbf{2}$\,\textbf{eV}\ \ \ \ \ \= W$^+$/H$^+$\,=\,\textbf{12}\\
 \ \ $n_\mathrm{e}=80$\,cm$^{-3}$ \> $T_\mathrm{e}=\mathbf{2}$\,\textbf{eV} \> W$^+$/H$^+$\,=\,\textbf{12}\\
 \ \ $n_\mathrm{e}=\textbf{60}$\,\textbf{cm}$^{\mathbf{-3}}$ \> $T_\mathrm{e}=1$\,eV \> W$^+$/H$^+$\,=\,\textbf{12}\\
 \ \ $n_\mathrm{e}=\textbf{60}$\,\textbf{cm}$^{\mathbf{-3}}$ \> $T_\mathrm{e}=3$\,eV \> W$^+$/H$^+$\,=\,\textbf{12}\\
 \ \ $n_\mathrm{e}=\textbf{60}$\,\textbf{cm}$^{\mathbf{-3}}$ \> $T_\mathrm{e}=\mathbf{2}$\,\textbf{eV} \> W$^+$/H$^+$\,=\,6\\
 \ \ $n_\mathrm{e}=\textbf{60}$\,\textbf{cm}$^{\mathbf{-3}}$ \> $T_\mathrm{e}=\mathbf{2}$\,\textbf{eV} \> W$^+$/H$^+$\,=\,30
\end{tabbing}
\noindent The numbers in bold indicate the values used thus far.  In each case, only $one$ of the three constraints is different from the original set ($n_\mathrm{e}=60$ cm$^{-3}$, $T_\mathrm{e}=2$ eV, W$^+$/H$^+$\,=\,12).  This modest sampling allows us to illustrate the dependence of plasma composition to each primary constraint.  

Figure \ref{barchart} shows the best fit for each case.  The parameter combination giving model results in best agreement with each set of the constraints is shown in the left panel of each bar chart.  In each chart, the original baseline case has been normalized to one, and the other two cases are given relative to this value (see Table \ref{nominalOutput} to determine the actual value for each parameter).  We have also generated sensitivity contours similar to those presented in Figures \ref{tehVfeh} and \ref{tauVfeh} for each of these new constraint combinations (not shown).  From the $f_\mathrm{diff}=0.05$ contours, we have derived the ranges over-plotted on the parameters in Figure \ref{barchart}.  No such ranges exist in the cases of $T_\mathrm{e}=1$ eV and $T_\mathrm{e}=3$ eV because the best fits have a total fractional difference of 0.43 and 0.19, respectively.

Water-group ion composition ratios are plotted in the right panel of each bar plot.  Densities and temperatures have not been shown because we are primarily interested in the effect on composition.  The top bar chart shows that neutral source rate, hot-electron temperature, and hot-electron fraction respond monotonically to $n_\mathrm{e}$.  In the middle chart, the neutral source and the hot-electron temperature decrease, while the hot-electron fraction and transport time increase with $T_\mathrm{e}$.  

Electron density imposes the weakest change in composition, and the W$^+$/H$^+$ ratio is the strongest driver.  In particular, H$_3$O$^+$ is the most abundant water-group ion only when W$^+$/H$^+$\,=\,30.  This is an important result because obtaining the W$^+$/H$^+$ ratio from the CAPS data is more straightforward than obtaining the separate, individual abundances of each water-group species \citep{wilson2008}.  Increasing the W$^+$/H$^+$ ratio requires a lower hot-electron temperature.  One reason for this is that the impact ionization rate of H depends less on the hot-electron temperature than do the water-group ionization rates (Figure \ref{WandHimpIonRates}).  Therefore, as the hot-electron temperature lowers, the W$^+$/H$^+$ ratio increases.  H$_3$O$^+$ on the other hand, is the only water-group ion that does not require hot-electron impact ionization to thrive.  In fact, it serves as a sink for both OH$^+$ and H$_2$O$^+$ ions via charge exchange (Table \ref{reactionSummary}).  Thus, as required by the fit in Figure \ref{barchart}, the hot-electron temp drops to raise the W$^+$/H$^+$ ratio and in the process increases the H$_3$O$^+$/W$^+$ ratio.

\section{Hot-electron ($f_\mathrm{eh}$) Modulation}\label{fehModulation}
\cite{steffl4} found that the interaction of two hot-electron populations orbiting Jupiter with System III and System IV periods are required to recover the Cassini UVIS temporal observations of the Io torus composition, both in terms of amplitude and rotational period.  \cite{2008GeoRL..3503107D} propose that the observed azimuthal electron density modulation at Saturn \citep{gurnett2007} is also caused by an azimuthally varying hot-electron abundance.  Motivated by their research, we present compositional sensitivity to a magnetic-longitude-dependent hot-electron fraction.  Figure \ref{fehVoutput} shows the response of the Enceladus torus to a prescribed sinusoidal hot-electron fraction:
\begin{equation}
\label{modulatedFeh}
f_\mathrm{eh}(\lambda_\mathrm{mag})=f_\mathrm{eh}^\mathrm{baseline}+0.0016\sin(\lambda_\mathrm{mag}),
\end{equation} 
where $f_\mathrm{eh}^\mathrm{baseline}$\,=\,0.0046 and $\lambda_\mathrm{mag}$ is magnetic longitude with arbitrary phase.  The modulation amplitude is chosen so that $n_\mathrm{e}$ modulates by roughly a factor of two ($\approx$\,40--80 cm$^{-3}$) according to the Saturnian kilometric radio emission analysis by \cite{gurnett2007} (see their Figure 2).  All other parameters are fixed at the baseline values listed in Table \ref{nominalOutput}.

The top panel of Figure \ref{fehVoutput} illustrates that $T_\mathrm{e}$ and $n_\mathrm{e}$ are in phase with $f_\mathrm{eh}$.  Impact dissociation and ionization by hot electrons (Table \ref{reactionSummary}) drive the W$^+$/H$^+$ ratio out of phase with $f_\mathrm{eh}$.  All quantities are plotted on the same logarithmic scale to show how linearly they respond to the hot-electron modulation.   

The middle panel shows the water-group compositional response to $f_\mathrm{eh}$ (plotted on the same logarithmic scale as the top panel).  The baseline composition ratios at $\lambda_\mathrm{mag}=0^\circ$ (and 360$^\circ$) are given in Table \ref{nominalOutput}.  The two hemispheres $0^\circ$--$180^\circ$ and $180^\circ$--$360^\circ$ are defined by a higher- and lower-than-baseline $f_\mathrm{eh}$, respectively.  As $f_\mathrm{eh}$ increases in the $0^\circ$--$180^\circ$ hemisphere, the hierarchy becomes $\mathrm{OH^+>H_2O^+>O^+>H_3O^+}$.  As $f_\mathrm{eh}$ drops in the $180^\circ -360^\circ$ hemisphere, H$_2$O$^+$ begins to dominate and O$^+$ becomes the minor species ($\mathrm{H_2O^+>H_3O^+>OH^+>O^+}$). 

The bottom panel illustrates how strongly and how linearly composition and thermal-electron parameters respond to $f_\mathrm{eh}$.  The plot shows the normalized change in each parameter with respect to variation in $f_\mathrm{eh}$.  Points are missing at the middle and endpoints because the plotted function diverges there as $\Delta f_\mathrm{eh}\equiv f_\mathrm{eh}-f_\mathrm{eh}^\mathrm{baseline} \rightarrow0$.  The solid and dashed lines respectively indicate quantities in and out of phase with $f_\mathrm{eh}$.  Quantities in phase with $f_\mathrm{eh}$ ($n_\mathrm{e}$, $T_\mathrm{e}$, O$^+$/W$^+$, and OH$^+$/W$^+$) have essentially flat curves, indicating linearity with $f_\mathrm{eh}$.  O$^+$/W$^+$ responds to perturbations 3--4 times less strongly than the rest.  Out-of-phase quantities (W$^+$/H$^+$, H$_2$O$^+$/W$^+$, and H$_3$O$^+$/W$^+$) also behave linearly with the exception of H$_2$O$^+$/W$^+$, which always reacts weakly and practically not at all when $f_\mathrm{eh}$ is low.  H$_3$O$^+$/W$^+$ and W$^+$/H$^+$ are always driven more efficiently than one-to-one.

Correlations and anti-correlations between electron density and composition such as those presented in Figure \ref{fehVoutput} have already been observed at Jupiter by \cite{steffl3}.  In particular, they find that S$^+$ is in phase, while S$^{3+}$ is 180$^\circ$ out-of-phase with equatorial electron density modulation in the Io torus. 

\section{Conclusions}\label{conclusion}
We have compared output from our model to the constraints on thermal-electron density ($n_\mathrm{e}$), thermal-electron temperature ($T_\mathrm{e}$), and mixing ratio of water-group ions to protons (W$^+/$H$^+$) by exploring the space spanned by the following four parameters:  neutral source rate ($N_\mathrm{src}$), hot-electron temperature ($T_\mathrm{eh}$), hot-electron density ($n_\mathrm{eh}\equiv f_\mathrm{eh}n_\mathrm{e}$), and radial transport time scale ($\tau_\mathrm{trans}$).

Our important results are:

\begin{enumerate}
 \item For the constraint choices of $n_\mathrm{e}=60\ \mathrm{cm}^{-3}$, $T_\mathrm{e}=2$ eV, and W$^+$/H$^+$=12, we find the following limits on the parameters:
\begin{enumerate}
  \item[\ ] $1.5\lesssim N_\mathrm{src}/(10^{-4}\mathrm{\ cm^{-3}\ s^{-1}})\lesssim 2.7$ 
  \item[\ ] $12\ \mathrm{days} \lesssim \tau_\mathrm{trans}$
  \item[\ ] $0.3\lesssim f_\mathrm{eh}(\%)\lesssim0.9$
  \item[\ ] $100\lesssim T_\mathrm{eh}/\mathrm{eV}\lesssim250$.
\end{enumerate}
\noindent  For a volume of $2\pi(4R_\mathrm{S})(2R_\mathrm{S})^2$, the source rate can be scaled to give a mass source rate of  $100\lesssim N_\mathrm{src}/(\mathrm{kg\ s^{-1}})\lesssim 180$.  We find that $f_\mathrm{H^+}$ (the fraction of protons confined to the equator) is strongly coupled to the hot-electron population and has not been constrained by this study (Section \ref{varyingProtonDilution}).
The solution space can be compared with future measurements of the parameters ($N_\mathrm{src},\ f_\mathrm{eh},\ T_\mathrm{eh},\ \tau_\mathrm{trans}$) and composition mixing ratios.  Upper limits on UV power emanating from the Enceladus torus from neutral oxygen at 1304, 1356, and 6300\,\AA\  would be very useful.  

  \item With the full water-based chemistry, photo- plus impact ionization is nearly as important as charge exchange at providing energy by way of fresh pickup ions (Figure \ref{enerFlow}).
  \item The water-based chemistry (particularly dissociative recombination) increases the neutral-to-ion ratio from 12 (the \cite{2007GeoRL..3409105D} oxygen-based model) to $\approx 40$.  Further, the Enceladus torus remains neutral-dominated even when the thermal-electron temperature approaches the temperature of electrons in Jupiter's Io torus (6 eV).
  \item The H$_3$O$^+$/W$^+$ ratio is directly correlated with the W$^+$/H$^+$ ratio (Figure \ref{barchart}), implying that H$_3$O$^+$ is strongly anti-correlated with H$^+$ (Section \ref{alternativeConstraints}).  This result is important because obtaining the W$^+$/H$^+$ ratio is more straightforward than obtaining individual water-group abundances from the CAPS data.  However, \cite{2008P&SS...56....3S} have shown, based on CAPS SOI data, that H$_3$O$^+$ dominates the water-group near the orbit of Enceladus (their Figure 15).  The dominance of H$_3$O$^+$ seen in the CAPS data has not been obtained by our model with the given constraints.  It is likely (manuscript in preparation) that the local interaction of the corotating plasma with the Enceladus plumes may contribute significantly to the  H$_3$O$^+$ abundance \citep{cravens2009}.
  \item Hot electrons are necessary in the Enceladus to enhance ionization torus but do not directly contribute more that 1\% of the total electron population.
  \item Significant variation in composition can be driven by a small perturbation in the hot-electron population (Figure \ref{fehVoutput}).  
\end{enumerate}
The sensitivity study presented here will be useful in a future interpretation of longitudinal and temporal observations (e.g., \cite{gurnett2007}) of the Enceladus torus, especially in the context of a modulating hot-electron density.  

%
%
%
%
%
%

%
%
%
%

\begin{acknowledgments}
This study was supported by NASA award number NNX08AB17G under the Cassini Data Analysis Program.  CHIANTI is a collaborative project involving the NRL (USA), RAL (UK), MSSL (UK), the Universities of Florence (Italy) and Cambridge (UK), and George Mason University (USA).  Generous allocations of computing time provided by the OU Supercomputing Center for Education and Research (OSCER) at the University of Oklahoma permitted the high-resolution grid calculation.
\end{acknowledgments}

%
%
%
%
%
%
%
%

%
%


%
%

\newpage
\begin{table}
  \centering 
 \begin{tabular}{ l @{\ \ : } l @{\ \ \ \ } l @{: } l  }
  \hline
\multicolumn{2}{c}{\rule{0pt}{2.6ex}Data (Constraints)}& \multicolumn{2}{c}{Baseline Fit}\\
  \hline
  \hline
 \rule{0pt}{2.6ex}$n_\mathrm{e}$/cm$^{-3}$&60 &$T_\mathrm{eh}/$eV&160\\ 
  $T_\mathrm{e}$/eV&2.0&$f_\mathrm{eh}$&0.46\,\%\\
   W$^+$/H$^+$&12&$f_{\mathrm{H}^+}$&1.0\\
  \multicolumn{2}{c}{} &$\tau_\mathrm{trans}$/days&26\\ 
\multicolumn{2}{c}{\rule[-2.2ex]{0pt}{0pt}}  &$N_\mathrm{src}$/cm$^{-3}$\,s$^{-1}$&2.0E-4 \\
\multicolumn{2}{c}{Neutral} &\multicolumn{2}{c}{\multirow{2}{*}{Mixing Ratios}}\\
\multicolumn{2}{c}{Densities (cm$^{-3}$)} &\multicolumn{2}{c}{ }\\
   \hline
\rule{0pt}{2.6ex}$n_\mathrm{H}$&720&O$^+$/W$^+$&0.15\\
$n_\mathrm{H_2}$&$\ll$ 1&OH$^+$/W$^+$&0.30\\
$n_\mathrm{O}$&700&H$_2$O$^+$/W$^+$&0.37\\
$n_\mathrm{OH}$&770&H$_3$O$^+$/W$^+$&0.18\\
$n_\mathrm{H_2O}$&190&O$^+$/H$^+$&1.8\\
\rule[-2.2ex]{0pt}{0pt}$n_\mathrm{H_3O}$&$-$&$\mathbf{W^+/H^+}$&\textbf{12}\\
\multicolumn{2}{c}{Ion/Electron}& \multicolumn{2}{c}{Ion/Electron}\\
\multicolumn{2}{c}{Densities (cm$^{-3}$)}& \multicolumn{2}{c}{Temperatures (eV)}\\
  \hline
\rule{0pt}{2.6ex}$\mathbf{n_\mathbf{e}}$&\textbf{60}&$\mathbf{T_\mathbf{e}}$&\textbf{2.0}\\
$n_\mathrm{eh}$&0.28&$T_\mathrm{eh}$&160\\
$n_\mathrm{H^+}$&4.6&$T_\mathrm{H^+}$&4.0\\
$n_\mathrm{H_2^+}$&$\ll$ 1&$T_\mathrm{H_2^+}$&6.5\\
$n_\mathrm{O^+}$&8.4&$T_\mathrm{O^+}$&38\\
$n_\mathrm{O^{++}}$&0.078&$T_\mathrm{O^{++}}$&35\\
$n_\mathrm{OH^+}$&17&$T_\mathrm{OH^+}$&39\\
$n_\mathrm{H_2O^+}$&20&$T_\mathrm{H_2O^+}$&42\\
\rule[-1.2ex]{0pt}{0pt}$n_\mathrm{H_3O^+}$&9.8&$T_\mathrm{H_3O^+}$&42\\
\hline
 \end{tabular} 
  \caption{Model constraints (see Section \ref{constraints} for references) and output for the best fit baseline solution.  The fit is defined as the combination of parameters $T_\mathrm{eh}$, $f_\mathrm{eh}$, $f_\mathrm{H^+}$, $\tau_\mathrm{trans}$, and $N_\mathrm{src}$ that gives the smallest total fractional difference between the data and the model output.  Notice that the model output for the constraints (in bold) agrees with the data to at least two significant figures.  No reaction leads to H$_3$O in our set of reactions (appendix).}
\label{nominalOutput}
\end{table}

\clearpage
\pagebreak
\newpage
\begin{table}
  \centering 
  \begin{tabular}{l  l@{\ } r @{$\times$}l  @{\ }r   @{$\times$}l l c|l l@{\ } r @{$\times$}l  @{\ }r   @{$\times$}l }
   \hline
\rule{0pt}{2.6ex} & \multicolumn{1}{c}{Mechanism}  & \multicolumn{2}{l}{Rate (s$^{-1}$)}&\multicolumn{2}{l}{$\tau$ (days)}& &&& \multicolumn{1}{c}{Mechanism}  & \multicolumn{2}{l}{Rate (s$^{-1}$)}&\multicolumn{2}{l}{$\tau$ (days)} \\
  \hline  
  \hline  
 \rule{0pt}{2.6ex}H &   Charge Exchange & 7.3&10$^{-8}$ & 1.6&10$^{2}$&&&H$^+$ &  Charge Exchange & 6.1&10$^{-6}$ & 1.9&10$^{0}$\\
&   Impact Ionization & 9.6&10$^{-9}$ & 1.2&10$^{3}$& &&  &Radial Transport & 4.4&10$^{-7}$ & 2.6&10$^{1}$\\
&   Photoionization & 8.0&10$^{-10}$ & 1.5&10$^{4}$& & & &Recombination & 5.1&10$^{-9}$ & 2.3&10$^{3}$\\
\hline
%
\rule{0pt}{2.6ex}H$_2$  &      Impact Dissociaton & 7.6&10$^{-7}$ & 1.5&10$^{1}$&&&H$_2^+$  &  Charge Exchange & 2.6&10$^{-6}$ & 4.4&10$^{0}$\\
& Charge Exchange & 5.2&10$^{-8}$ & 2.2&10$^{2}$&& &&Dissociative Recomb. & 1.2&10$^{-6}$ & 9.6&10$^{0}$\\
  &  Photoionization & 6.9&10$^{-10}$ & 1.7&10$^{4}$& & &&Radial Transport & 4.4&10$^{-7}$ & 2.6&10$^{1}$\\
  &  Photodissociation & 4.9&10$^{-10}$ & 2.4&10$^{4}$&&&\multicolumn{6}{c}{} \\
\hline
\rule{0pt}{2.6ex}O&     Charge Exchange & 6.6&10$^{-8}$ & 1.7&10$^{2}$&&&O$^+$ &  Charge Exchange & 3.1&10$^{-6}$ & 3.7&10$^{0}$\\
 &  Impact Ionization & 2.6&10$^{-8}$ & 4.5&10$^{2}$&&&& Radial Transport & 4.4&10$^{-7}$ & 2.6&10$^{1}$\\
&   Photoionization & 2.3&10$^{-9}$ & 5.0&10$^{3}$&&&&  Impact Ionization & 7.5&10$^{-9}$ & 1.5&10$^{3}$\\
\multicolumn{6}{c}{}  &&&& Recombination & 1.9&10$^{-11}$ & 6.1&10$^{5}$\\
\multicolumn{6}{c}{}\rule{0pt}{2.6ex}&&&O$^{++}$   & Radial Transport & 4.4&10$^{-7}$ & 2.6&10$^{1}$\\
\multicolumn{6}{c}{}&&&&  Charge Exchange & 3.7&10$^{-7}$ & 3.2&10$^{1}$\\
\multicolumn{6}{c}{}&&&&   Recombination & 1.1&10$^{-10}$ & 1.0&10$^{5}$\\
\hline
%
 \rule{0pt}{2.6ex}OH&     Photodissociation & 5.5&10$^{-8}$ & 2.1&10$^{2}$&&&OH$^+$\  & Charge Exchange & 1.1&10$^{-6}$ & 1.1&10$^{1}$ \\
&  Impact Ionization & 3.5&10$^{-8}$ & 3.3&10$^{2}$ &&& &    Dissociative Recomb. & 5.8&10$^{-7}$ & 2.0&10$^{1}$\\
 &  Impact Dissociation & 2.7&10$^{-8}$ & 4.3&10$^{2}$& &&& Radial Transport & 4.4&10$^{-7}$ & 2.6&10$^{1}$\\
 &   Charge Exchange & 1.6&10$^{-8}$ & 7.4&10$^{2}$&&&\multicolumn{6}{c}{}\\
  &  Photoionization & 3.7&10$^{-9}$ & 3.1&10$^{3}$&&&\multicolumn{6}{c}{}\\
\hline
\rule{0pt}{2.6ex}H$_2$O&    Impact Dissociation & 4.9&10$^{-7}$ & 2.4&10$^{1}$&&&H$_2$O$^+$ &  Dissociative Recomb. & 1.2&10$^{-6}$ & 9.9&10$^{0}$\\
&   Charge Exchange & 3.6&10$^{-7}$ & 3.2&10$^{1}$& &&  & Radial Transport & 4.4&10$^{-7}$ & 2.6&10$^{1}$\\
 &    Photodissociation & 1.3&10$^{-7}$ & 9.3&10$^{1}$&&& &Charge Exchange & 4.0&10$^{-7}$ & 2.9&10$^{1}$\\
&Impact Ionization& 5.2&10$^{-8}$ & 2.2&10$^{2}$&&&\multicolumn{6}{c}{}\\
&     Photoionization & 4.5&10$^{-9}$ & 2.6&10$^{3}$&&&\multicolumn{6}{c}{}\\
\hline
\multicolumn{6}{c}{}&&&\rule{0pt}{2.6ex}H$_3$O$^+$    & Dissociative Recomb. & 8.1&10$^{-7}$ & 1.4&10$^{1}$\\
 \multicolumn{6}{c}{} &&&&  Radial Transport & 4.4&10$^{-7}$ & 2.6&10$^{1}$\\
\hline
\end{tabular} 
  \caption{Baseline lifetimes for each species by mechanism in descending order of frequency (Rate\,$=$\,$1/\tau$).  Electron-impact ionization and photoionization include processes that are both ionizing and dissociative.  A listing of lifetimes by reaction can be found in the appendix.}
  \label{time scalesMech}
\end{table}

\clearpage
\pagebreak
\newpage
\begin{table}
 \centering 
  \begin{tabular}{l @{ $\rightarrow$ } l |l @{ $\rightarrow$ }l}
   \hline
\multicolumn{4}{c}{\rule{0pt}{2.6ex}Dominant Reactions}\\
  \hline 
  \hline 
\rule{0pt}{2.6ex}H   + e$_\mathrm{h}$& H$^+$ + 2e   &  H$^+$ + H & H + H$^+$\\
O   + e$_\mathrm{h}$   & O$^+$ + 2e&  H$^+$ + O & H + O$^+$\\ 
OH   + e    &OH$^+$  + 2e&  H$^+$ + OH & H + OH$^+$\\
OH   + e$_\mathrm{h}$    & OH$^+$  + 2e&  H$^+$ + H$_2$O & H + H$_2$O$^+$\\
H$_2$O  + e$_\mathrm{h}$   & H$_2$O$^+$ + 2e&  O$^+$ + H & O + H$^+$\\ 
H$_2$O  + e$_\mathrm{h}$    & OH$^+$  + H + 2e&  O$^+$ + O & O + O$^+$\\ 
H$_2$O  + e$_\mathrm{h}$    & H$^+$   + OH + 2e &  O$^+$ + OH & O + OH$^+$\\ \cline{1-2}
\rule{0pt}{2.0ex}OH + e & O + H + e&  O$^+$ + H$_2$O & O + H$_2$O$^+$\\
OH + e$_\mathrm{h}$ & O + H + e&  OH$^+$ + OH & O + H$_2$O$^+$\\
H$_2$O + e & OH + H + e& OH$^+$ + H$_2$O & OH + H$_2$O$^+$\\ 
H$_2$O + e$_\mathrm{h}$ & OH + H + e& OH$^+$ + H$_2$O & O + H$_3$O$^+$\\ \cline{1-2}
\rule{0pt}{2.0ex}O + $\gamma$ & O$^+$   + e& H$_2$O$^+$ + H$_2$O & OH + H$_3$O$^+$\\
OH + $\gamma$ & OH$^+$  + e& H$_2$O$^+$ + H$_2$O & H$_2$O + H$_2$O$^+$\\ \cline{3-4}
\rule{0pt}{2.0ex}OH + $\gamma$ & O + H&OH$^+$  + e   & O   + H\\
H$_2$O + $\gamma$ & H + OH&H$_2$O$^+$ + e   & OH  + H\\
H$_2$O + $\gamma$ & H$_2$   + O  &H$_3$O$^+$ + e   & OH  + H$_2$\\
\hline
\end{tabular} 
  \caption{List of the most important reactions for the baseline case.  The left column gives the relevant impact, dissociative, and photolytic reactions, and the right column gives all relevant charge exchanges and recombinations.   The full set of reactions are given in the electronic appendix, but steady-state densities and temperatures are all within 3\% of the properly calculated values when only the above reactions are turned on.}
\label{reactionSummary}
\end{table}


\clearpage
\pagebreak
\newpage
\begin{figure}
\centering
\noindent\includegraphics[width=5.5in,angle=0]{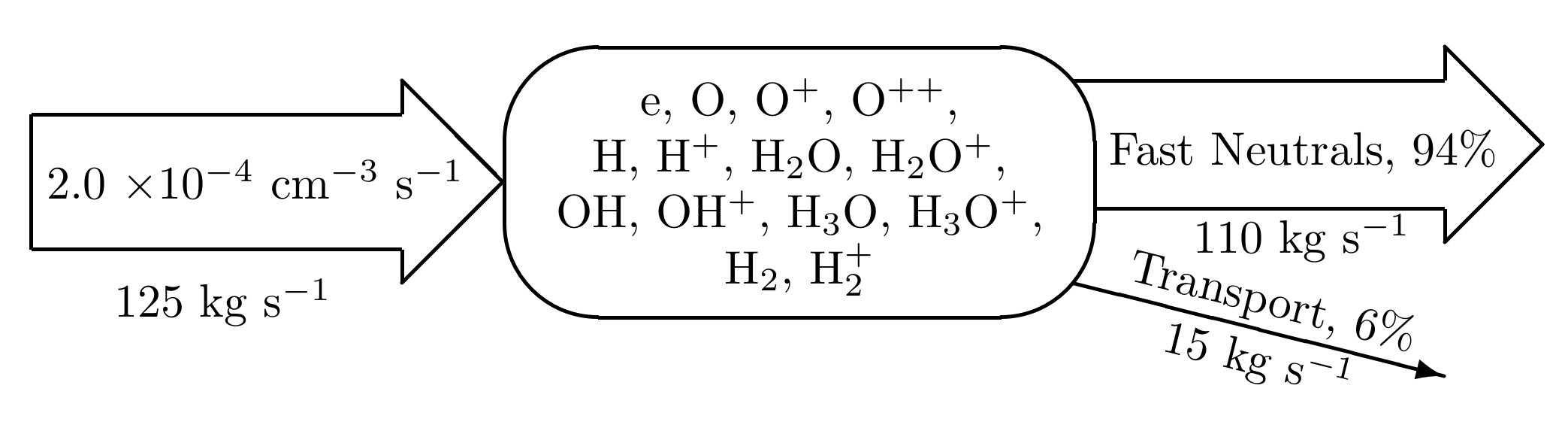}
 \caption{Particle flow for the baseline case.  We have assumed a torus volume of $2\pi(4R_\mathrm{S})(2R_\mathrm{S})^2$ to calculate the volumetric mass flow.  The percentages given here are for particle $number$ (not mass).  Individual species contributions to the particle outflow can be found in the electronic appendix.}
 \label{partFlow}
\end{figure}

 \begin{figure}
\centering
\noindent\includegraphics[width=5.5in,angle=0]{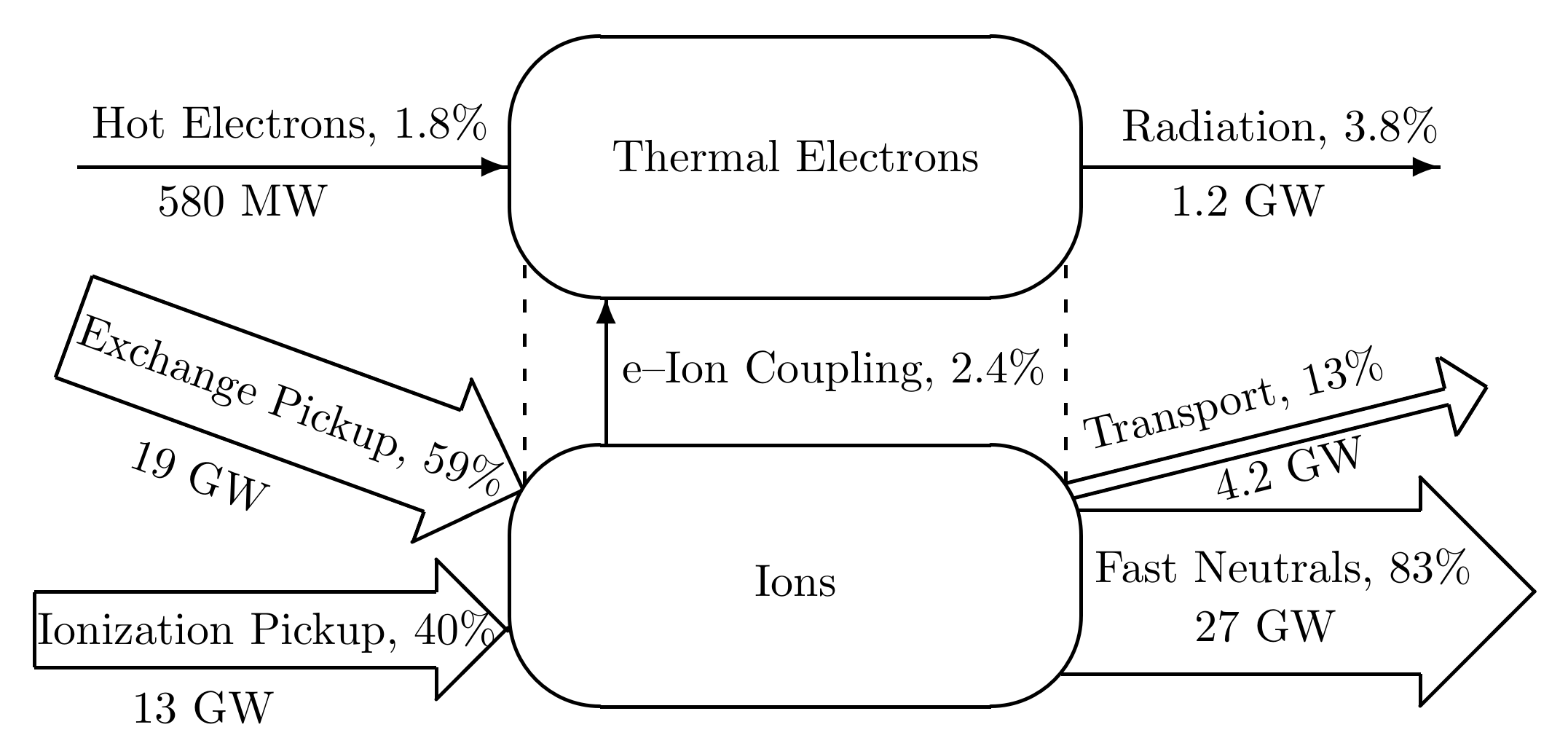}
 \caption{Energy flow for the baseline case.  We have assumed a torus volume of $2\pi(4R_\mathrm{S})(2R_\mathrm{S})^2$ to calculate the volumetric energy flow.  Individual species contributions to the energy outflow can be found in the electronic appendix.}
 \label{enerFlow}
\end{figure}
\clearpage
\pagebreak
\newpage

 \begin{figure}
\centering
\noindent\includegraphics[width=6.5in,angle=0]{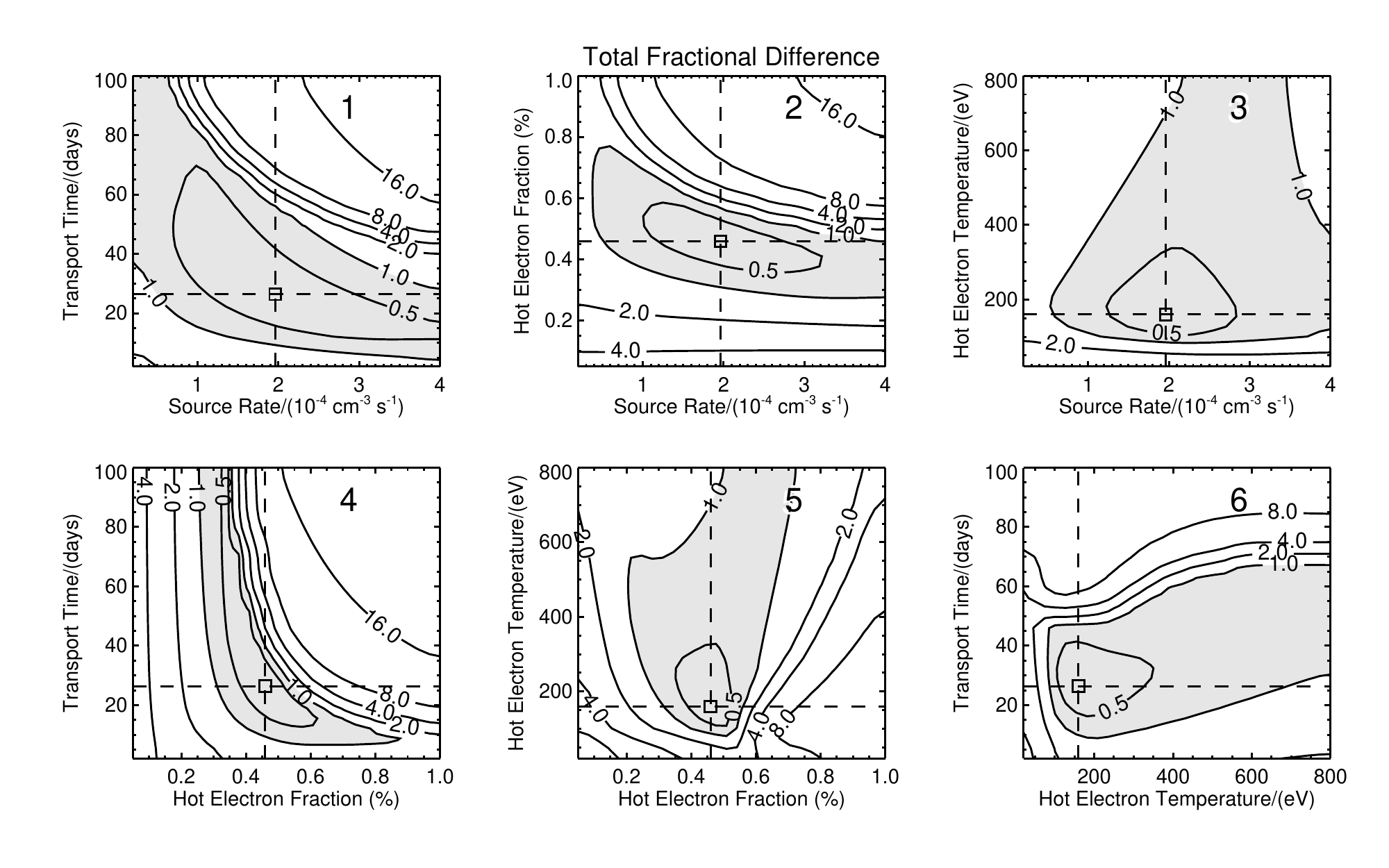}
 \caption{Sensitivity plots of $f_\mathrm{diff}$ for every parameter combination.  In each case, the remaining three parameters are fixed at their baseline values (Table \ref{nominalOutput}) to ascertain trends due solely to variation of a single parameter at a time.  The intersection of dashed lines indicates the baseline solution, and the gray shading inside of $f_\mathrm{diff}=1$ is intended to guide the eye.  (Contours are plotted logarithmically.)}
 \label{sensPlot}
 \end{figure}
\clearpage
\pagebreak
\newpage

\begin{figure}
\centering
\noindent\includegraphics[width=5.in,angle=0]{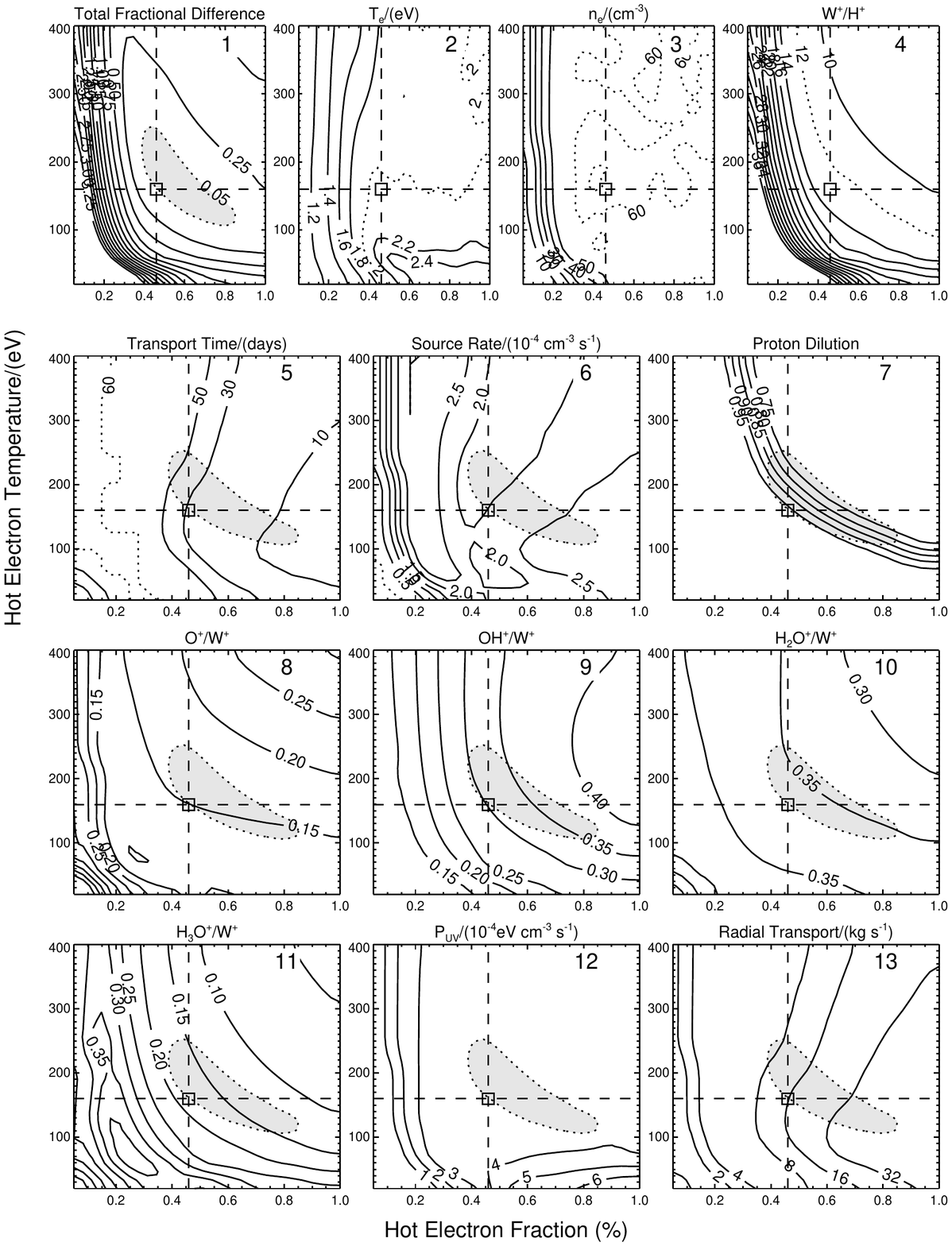}
 \caption{Sensitivity plots between hot-electron temperature ($T_\mathrm{eh}$) and hot-electron fraction ($f_\mathrm{eh}$) generated from the grid search.  The over-plotted box represents the baseline solution, discussed in Section \ref{BaselineSolution}.  The $f_\mathrm{diff}=0.05$ curve (Panel 1) has been over-plotted on Panels 5--13 in gray.  The three parameters in Panels 5--7 ($\tau_\mathrm{trans},\ N_\mathrm{src},\ \mathrm{and}\ f_\mathrm{H^+}$) have taken on values yielding best agreement between model output and $n_\mathrm{e}$\,=\,60 cm$^{-3}$, $T_\mathrm{e}$\,=\,2 eV, and W$^+$/H$^+$\,=\,12.  Combinations of $T_\mathrm{eh}$ and $f_\mathrm{eh}$ within the gray contour are consistent with these constraints.  All panels are discussed in Section \ref{gridSearchResults}.}
 \label{tehVfeh}
\end{figure}
\clearpage
\pagebreak
\newpage

\begin{figure}
\centering
\noindent\includegraphics[width=5.in,angle=0]{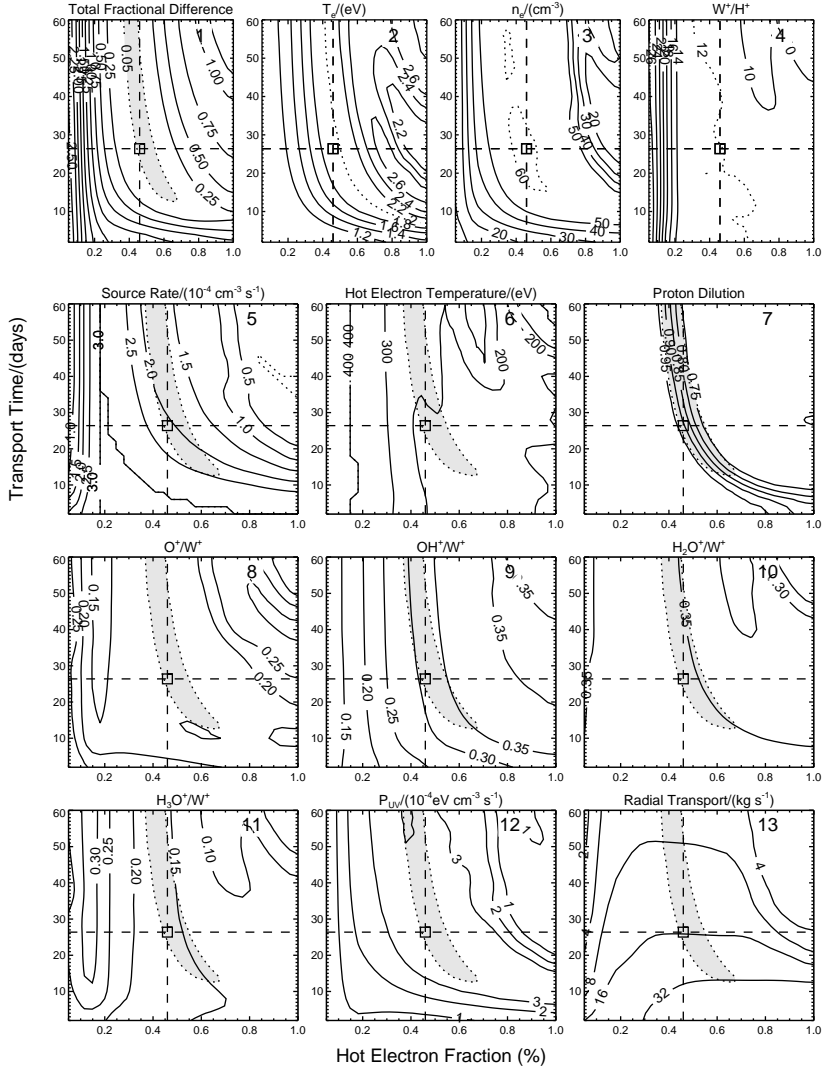}
\caption{Sensitivity plots between radial transport time scale ($\tau_\mathrm{trans}$) and hot-electron fraction ($f_\mathrm{eh}$) generated from the grid search.  The over-plotted box represents the baseline solution, discussed in Section \ref{BaselineSolution}.  The $f_\mathrm{diff}=0.05$ curve (Panel 1) has been over-plotted on Panels 5--13 in gray.  The three parameters in Panels 5--7 ($N_\mathrm{src},\ \tau_\mathrm{trans},\ \mathrm{and}\ f_\mathrm{H^+}$) have taken on values yielding best agreement between model output and $n_\mathrm{e}$\,=\,60 cm$^{-3}$, $T_\mathrm{e}$\,=\,2 eV, and W$^+$/H$^+$\,=\,12.  Combinations of $\tau_\mathrm{trans}$ and $f_\mathrm{eh}$ within the gray contour are consistent with these constraints.  All panels are discussed in Section \ref{gridSearchResults}.}
 \label{tauVfeh}
\end{figure}

\clearpage
\pagebreak
\newpage
\begin{figure}
\centering
\noindent\includegraphics[width=6.5in,angle=0]{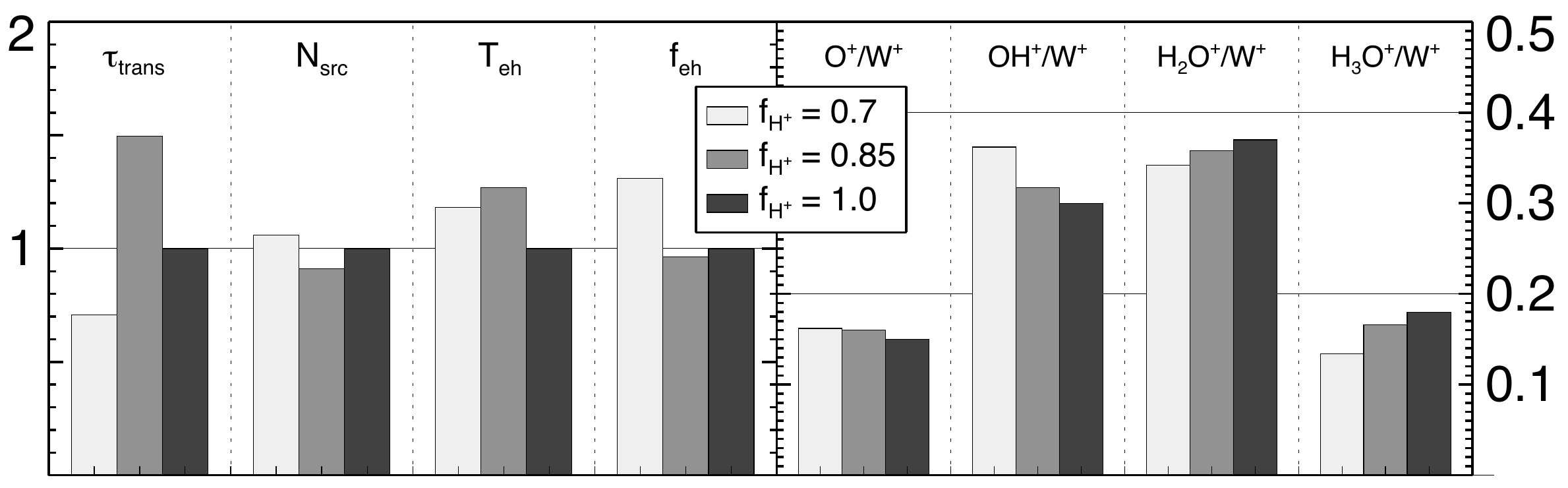}
 \caption{Solutions for various proton dilution factors ($f_\mathrm{H^+}=0.7,0.85,1.0$) found from the grid search in Section \ref{varyingProtonDilution}.  The parameters in the left panel are normalized to the values listed in Table \ref{nominalOutput}.  In the right panel we show the model output for water-group composition.  Proton production is strongly controlled by impact ionization of hydrogen by hot electrons, so $f_\mathrm{H^+}$ is coupled to $T_\mathrm{eh}$ and $f_\mathrm{eh}$.  This coupling diminishes the significance of $f_\mathrm{H^+}$ and effectively reduces the parameter space to four dimensions.}
 \label{fhpVoutput}
\end{figure}

\clearpage
\pagebreak
\newpage
\centering
\begin{figure}
\noindent\includegraphics[width=6.in,angle=0]{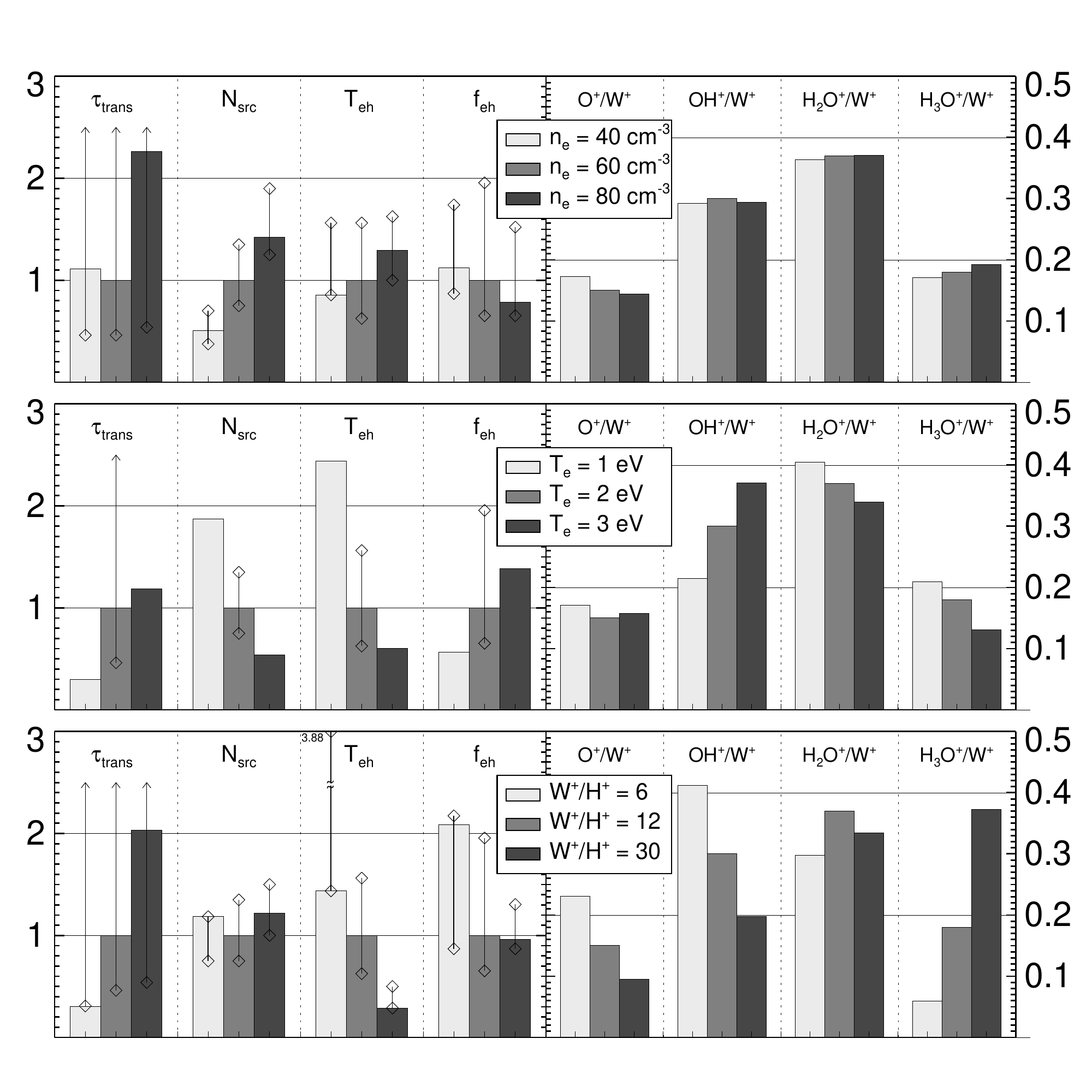}
 \caption{Solutions for primary constraint values representing a wide range of observations found from the grid search in Section \ref{alternativeConstraints}.  In each case, $n_\mathrm{e}=60$ cm$^{-3}$, $T_\mathrm{e}=2$ eV, and W$^+$/H$^+$ = 12 unless otherwise specified.  The parameter fits given in the left panel of each bar chart are normalized to the fit given in Table \ref{nominalOutput}.  The over-plotted ranges on the parameters are derived from the corresponding $f_\mathrm{diff}=0.05$ sensitivity contours.  No such ranges exist for the $T_\mathrm{e}=1$ eV and $T_\mathrm{e}=3$ eV solutions because the best fits have a fractional difference of 0.43 and 0.19, respectively.  The corresponding model output for water-group composition is presented on the right.  The choice of electron density ($n_\mathrm{e}$) has a weak effect, while the ratio W$^+$/H$^+$ ratio alters composition markedly.}
 \label{barchart}
\end{figure}

\clearpage
\pagebreak
\newpage
\begin{figure}
\centering
\noindent\includegraphics[width=4.5in,angle=0]{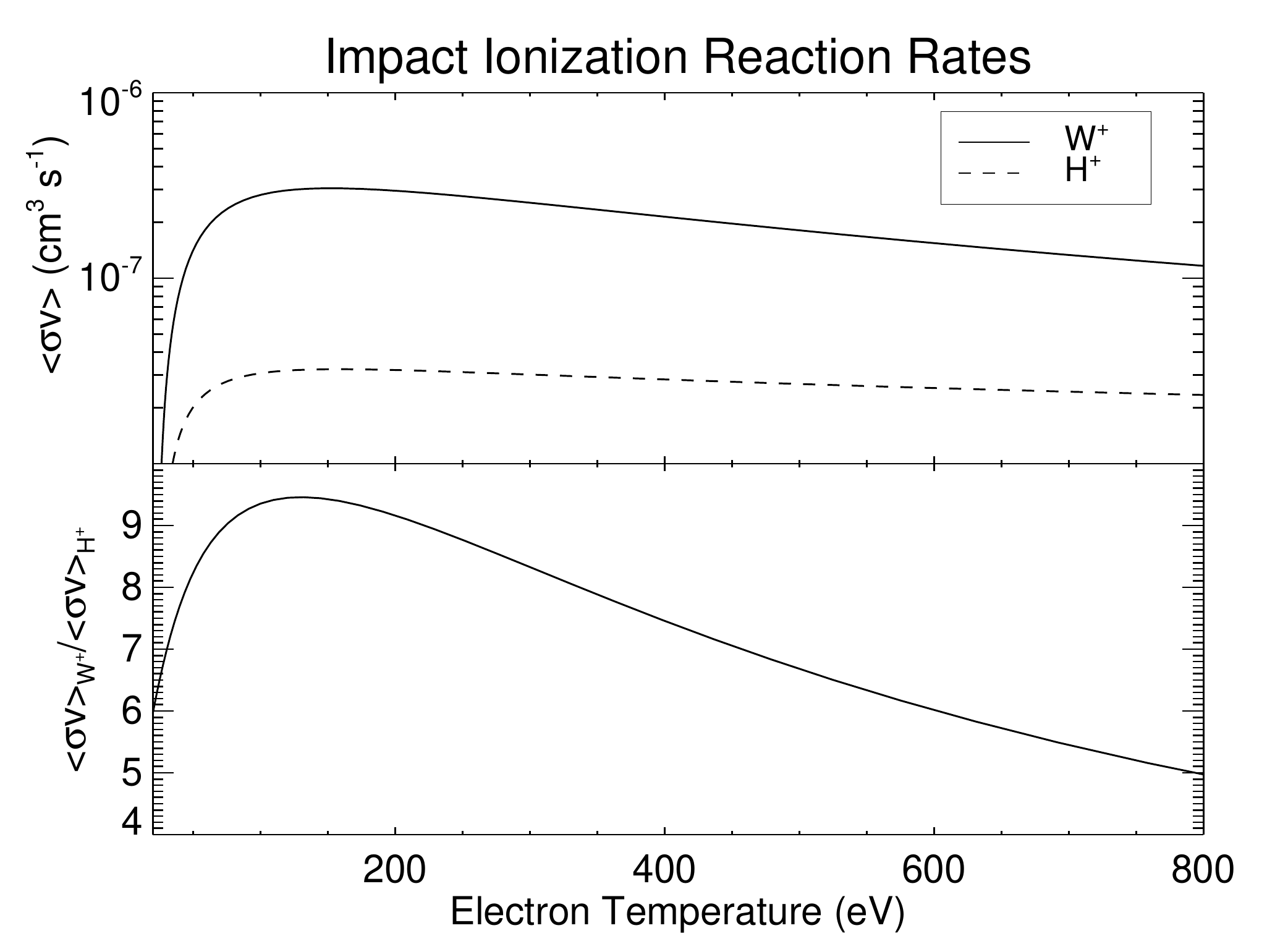}
 \caption{Electron-impact reaction rates for the water group (O\,+\,OH\,+\,H$_2$O) and hydrogen.  The bottom panel emphasizes that water-group reaction rates fall faster than hydrogen reaction rates as $T_\mathrm{eh}$ increases.}
 \label{WandHimpIonRates}
\end{figure}

\clearpage
\pagebreak
\newpage
\begin{figure}
\centering
\noindent\includegraphics[width=3.5in,angle=0]{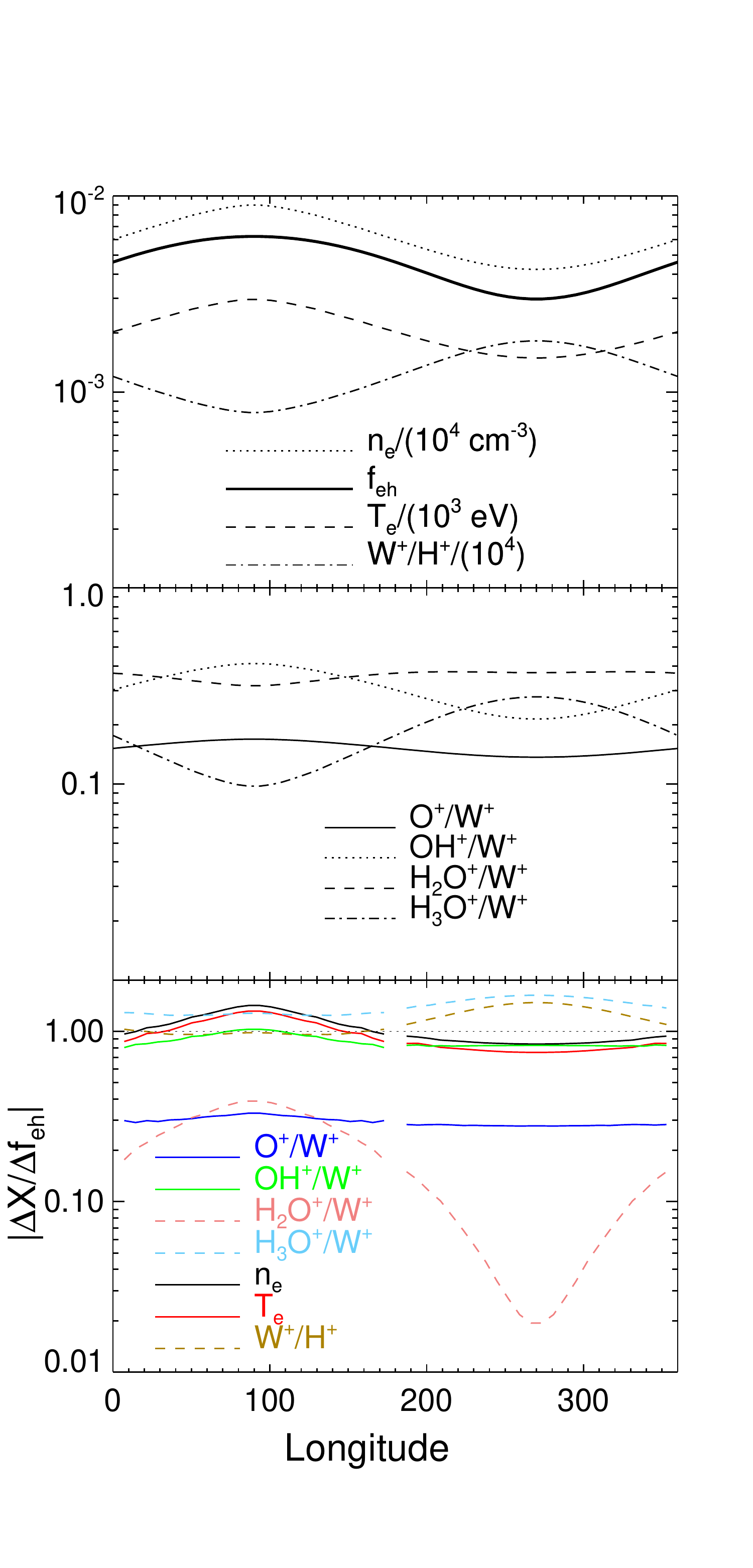}
 \caption{Composition variability due to hot-electron modulation.  The hot-electron fraction, $f_\mathrm{eh}$, alone drives the variation since all other parameters are held at their baseline values.  The quantities O$^+$/W$^+$, OH$^+$/W$^+$, $n_\mathrm{e}$, and $T_\mathrm{e}$ are in phase while W$^+$/H$^+$, H$_2$O$^+$/W$^+$, and H$_3$O$^+$/W$^+$ are out of phase with $f_\mathrm{eh}$.  The bottom panel shows how strongly and how linearly each quantity responds to perturbations in $f_\mathrm{eh}$; the solid lines represent quantities in phase with $f_\mathrm{eh}$, and the dashed lines represent quantities out of phase with $f_\mathrm{eh}$.}
 \label{fehVoutput}
\end{figure}

\end{article}



%
%
%
%
%
%

\clearpage
\pagebreak
\newpage

%
%
%
%
 

\end{document}